\documentclass[usenatbib]{mn2e}
\usepackage{fleqn,epsfig,txfonts}

\catcode`\@=11
\def\gta{\ifmmode{\,\mathrel{\mathpalette\@versim>\,}}
    \else{$\,\mathrel{\mathpalette\@versim>}\,$}\fi}
\def\lta{\ifmmode{\,\mathrel{\mathpalette\@versim<\,}}
    \else{$\,\mathrel{\mathpalette\@versim<}\,$}\fi}
\def\@versim#1#2{\lower 2.9truept \vbox{\baselineskip 0pt \lineskip
    0.5truept \ialign{$\m@th#1\hfil##\hfil$\crcr#2\crcr\sim\crcr}}}
\catcode`\@=12
\def\fracj#1#2{{\textstyle{#1\over#2}}}
\def\vx{{\bf x}}\def\vR{{\bf R}}\def\vz{{\bf z}}
\let\boldgrk=\gkvecten
\let\boldgrksc=\gkvecseven

\def\gkthing#1{{\mathchoice%
	{\hbox{{\boldgrk\char#1}}}
	{\hbox{{\boldgrk\char#1}}}
	{\hbox{{\boldgrksc\char#1}}}
	{\hbox{{\boldgrksc\char#1}}}}}
\def\vphi{\gkthing{30}}

\def\df{{\sc df}}
\def\AA{{\sc aa}}
\def\Rc{R_{\rm c}}\def\Rg{{R_{\rm g}}}
\def\zm{z_{\rm m}}
\def\ex#1{\langle#1\rangle}
\def\vc{v_{\rm c}}

\def\Phieff{\Phi_{\rm eff}}
\def\Phiad{\Phi_{\rm ad}}

\def\kms{\,{\rm km}\,{\rm s}^{-1}}

\def\pc{\,{\rm pc}}
\def\kpc{\,{\rm kpc}}

\def\e{{\rm e}}
\def\d{{\rm d}}
\def\msun{\,{\rm M}_\odot}

\def\figref#1{Fig.~\ref{#1}}

\renewcommand{\[}{\begin{equation}}
\renewcommand{\]}{\end{equation}}

\title[A new formula for fitting the azimuthal component of disc kinematics]
{A new formula for disc kinematics}

\author[Sch\"onrich \& Binney]{Ralph Sch\"onrich$^1$\thanks{E-mail:rasch@mpa-garching.mpg.del}, James Binney$^2$\\
 $^{1}$ Max Planck Institute for Astrophysics, Karl-Schwarzschild-Str. 1, D-85741 Garching \\
$^{2}$ University of Oxford, Rudolf-Peierls Centre for Theoretical Physics, Keble Road, Oxford OX1 3NP\\}

\begin{document}

\date{Draft, Sept 1, 2011}

\pagerange{\pageref{firstpage}--\pageref{lastpage}} \pubyear{2010}

\maketitle

\label{firstpage}

\begin{abstract}
In a disc galaxy the distribution of azimuthal components of velocity is very
skew. In the past this skewness has been modelled by superposed Gaussians. We
use dynamical arguments to derive an analytic formula that can be fitted to
observed velocity distributions, and validate it by fits to the velocities
derived from a dynamically rigorous model, and to a sample of local stars with
accurate space velocities. Our formula is much easier to use than a full
distribution function. It has fewer parameters than a multi-Gaussian fit, and
the best-fitting model parameters give insight into the underlying disc
dynamics. In particular, once the azimuthal velocities of a sample have been
successfully fitted, the apparatus provides a prediction for the
corresponding distribution of radial velocities $v_R$. An effective formula
like ours is invaluable when fitting to data for stars at some distance from
the Sun because it enables one to make proper allowance for the errors in
distance and proper motion when determining the underlying disc kinematics.
The derivation of our formula elucidates the way the
horizontal and vertical motions  are closely intertwined, and makes it
evident that no stellar population can have a scale height and vertical
velocity dispersions that are simultaneously independent of radius.
We show that the oscillation of a star perpendicular to the Galactic plane
modifies the effective potential in which the star moves radially in such a
way that the more vertical energy a star has, the larger is the mean radius
of its orbit. 
\end{abstract}


\section{Introduction}

Currently considerable effort is being invested in surveys of the solar
neighbourhood. Fifteen years ago the study of nearby stars was
revived by the Hipparcos mission, which pioneered space astrometry. Hipparcos
put ground-based astrometry onto a more secure foundation, so now useful
proper motions are available for tens of millions of stars. In the last
decade the proper motions have been complemented by photometric
surveys, both in the infrared and to fainter magnitudes at optical
wavelengths.  Finally, the RAVE \citep{RAVEI} and SEGUE \citep[see][]{york00,
Yanny09} surveys have measured nearly a million 
line-of-sight velocities. 

As a result of these major observational programmes, it is becoming possible
to determine the velocity distribution within the disc of our Galaxy, not
only at the location of the Sun, but also at significant distances,
especially at higher Galactic latitudes. Naturally one wants to quantify the
velocity distribution observed at some location $\vx$ in the Galaxy in an
efficient way. Conventionally one does this by imagining that the density of
stars in velocity space forms a ``velocity ellipsoid'' -- a triaxial
ellipsoidal region of over-density in velocity space. If the Galaxy were
axisymmetric (which is a reasonable first approximation), we would expect
that in the Galactic plane the principal axes of the the velocity ellipsoid
would be aligned with the coordinate directions of cylindrical polar
coordinates, $(R,z,\phi)$. As one moves above the plane, two of the
principal axes of the velocity ellipsoid are expected to tip slightly with
respect to the $\hat\vR$ and $\hat\vz$ directions. Let the components of
velocity parallel to these principal axes be denoted $v_1$ and $v_2$, where
$v_1\to v_R$ and $v_2\to v_z$ as $z\to0$. The third axis is expected to remain aligned with
the $\hat\vphi$ direction.

The distributions of the $v_1$ and $v_2$ components of velocity are expected
to be roughly Gaussian with vanishing means and to be to good
approximation aligned with the Galactic polar coordinates \citep[minor vertex
deviations as found in the solar neighbourhood by][will not be discussed
here]{Dehnen98}. Consequently, they can be characterised by
their standard deviations $\sigma_1$ and $\sigma_2$. The distribution of the
$v_\phi$ components peaks at a value of $v_\phi$ that is slightly smaller
than the circular speed $v_c$. However, it is not at all well modelled by a
Gaussian, because it is very skew, with many more stars at $v_\phi=v_c-v$
than at $v_c+v$, causing the population to have non-zero asymmetric drift.
Notwithstanding this skewness that was already known to Gustav Str\"omberg
\citep[][]{Stromberg27}, $v_\phi$ distributions have traditionally been
characterised by a mean and a standard deviation. Since a single Gaussian
fits the data very poorly, the observed distribution is frequently modelled
by a superposition of two Gaussians: then the overall distribution is
characterised by two means, two dispersions and the ratio of the numbers of
stars accounted for by each Gaussian, a total of five shape parameters.

The purpose of this note is to introduce a new representation of $v_\phi$
distributions that is more effective in the sense that it fits typical data
more accurately with fewer and physically more meaningful parameters.
Moreover, the new representation has a dynamical basis, so it is able to
connect the skewness of the $v_\phi$ distributions to the standard deviations
in $v_1$ and $v_2$. With the new representation, a single free shape
parameter suffices to describe the distribution in $v_\phi$ for the whole
population of stars in the solar cylinder, and two parameters are sufficient
to fit the distribution in $v_\phi$ of stars that have a given distance from
the plane.  The formula predicts in a natural way both the magnitude of the
asymmetric drift and the offset of the modal azimuthal velocity from the
circular velocity, neither of which is achieved by multiple Gaussians.

The paper is organised as follows. In Section \ref{sec:2d} we derive an
approximation to the $v_\phi$ distribution in an annulus in the Galactic disc
that can be used as standard for extragalactic measurements. As most samples
of the Galaxy are centred on certain Galactic altitudes $|z|$, in Section
\ref{sec:3d} we use the adiabatic approximation \citep{B10} to take into
account the vertical motions of stars. In \ref{sec:getf} we derive a formula
that accounts for the variation in the $v_\phi$ distribution with $|z|$ and
test it against the velocity distributions of more elaborate models. In
Section \ref{sec:backreact} we derive a formula for the way in which the
in-plane motion of a star depends on the extent of its excursions
perpendicular to the plane. The outcome is a small correction to the $v_\phi$
distribution derived in Section \ref{sec:getf}.  In Section \ref{sec:moments}
we give formulae from which the distributions of $v_R$ and $v_z$ follow once
the distribution in $v_\phi$ has been fitted. In Section \ref{sec:apptorus}
we show that our formulae provide good fits to the disc model of Binney \&
McMillan (2011; hereafter BM11), which has a rigorous
dynamical basis.  In Section \ref{sec:GCS} we demonstrate the practical
application of the formula by fitting data from the Geneva-Copenhagen Survey.
Section \ref{sec:sum} sums up and looks to the future.

\section{Velocity distribution as a 2D problem}\label{sec:2d}

There are three reasons for the asymmetry of the distribution of $v_\phi$
components of nearby stars: stars at low $v_\phi$ are approaching apocentre,
so they have guiding-centre radii $\Rg$ smaller than the solar radius, $R_0$.
As one moves inwards through the disc, not only does the density of stars
increase rapidly on account of the exponential increase in the surface
density $\Sigma(R)\propto\exp(-R/R_\d)$, but the random velocities of stars
also increase, so a greater fraction of all stars are on eccentric orbits
that carry them far from their guiding-centre radius $\Rg$. Moreover, the
effective potential in which a star oscillates around $\Rg$ rises much more
steeply at $R<\Rg$ than it does at $R>\Rg$, so stars spend more time beyond
$\Rg$ than they do interior to it. In fact, as a population of stars heats up
over its lifetime, the asymmetry of the effective potential causes the
population to expand spatially, and by conservation of angular momentum its
mean rotation rate diminishes. For all these reasons, there are many more
visitors reaching $R_0$ with guiding centres at $R_0-\Delta$ than at
$R_0+\Delta$.  A
functional form for $n(v_\phi)$ that is successful in fitting observed
distributions will reflect these facts.

Following \cite{Shu69} we decompose the energy of a disc star into three
parts. If the star were on a circular orbit with angular momentum $L_z$, it
would have energy
 \begin{equation}
E_c(L_z)=\Phieff(\Rg, L_z),
\end{equation}
 where with the Galactic potential in the plane $\Phi(R)$
\begin{equation}
\Phieff(R,L_z)\equiv{L_z^2\over2R^2}+\Phi(R)
\end{equation}
 and the guiding-centre radius $\Rg(L_z)$ solves the equation
$\Rg\vc(\Rg)=L_z$, with $\vc(R)$ the circular speed.  In addition to this
energy, the star has two smaller energies, namely the energy $E_z$ of
vertical motion and the energy $E_R$ of random motion within the plane. We
postpone discussion of $E_z$ and focus for now on stars with $E_z=0$, which
move in the plane. We have
 \begin{eqnarray}\label{eq:energy}
E_R&=&\fracj12v_R^2+\Phieff(R,L_z)-\Phieff(\Rg,L_z)\nonumber\\
&=&\fracj12v_R^2+\Delta\Phieff(R,L_z)
\end{eqnarray}
 where 
\begin{equation}\label{eq:defsDphi}
\Delta\Phieff(R,L_z)\equiv\Phieff(R,L_z)-\Phieff(\Rg,L_z).
\end{equation}

 Suppose that the disc's distribution function (\df) is \citep{Shu69} 
 \[\label{eq:defsf}
f(E_R,L_z)={F\over\sigma^2}\e^{-E_R/\sigma^2},
\]
 where $F(\Rg)$ is a function that determines the surface density of the
young disc and $\sigma(\Rg)$ is a function that determines how the radial
velocity dispersion varies with $R$. On account of the tendency noted above for a population
to expand radially as it heats up, if $F(\Rg)$ is the same for both cool and
hot populations, the hotter populations will have slightly larger radial
scale-lengths than the cool ones. Note that $\sigma$ gives the
intrinsic dispersion of stars at their guiding centre radius $\Rg$. The
dispersion $\langle v_R^2\rangle^{1/2}$ actually measured at some radius $R$
will have contributions from all populations that reach this radius and turns
out to be $\sim$10 per cent higher than the value of $\sigma$ for the stars
that have $\Rg=R$ on account of the presence of stars that have guiding
centres at $\Rg<R$. 

\cite{SBI} show that the
probability per unit area that a star with angular momentum $L_z$ will be
found at $R$ is
 \[\label{eq:givesP}
P(R|L_z)={K\over\sigma
R}\exp\left[-{\Delta\Phieff(R,L_z)\over\sigma^2}\right],
\]
 where $K(\Rg)$ is chosen such that $1=2\pi\int\d R\,RP$.

Let $n(v_\phi,R)\,\d v_\phi$ be the number per unit area of stars at $R$ with
$v_\phi$ in $(v_\phi,v_\phi+\d v_\phi)$. Then
 \[\label{eq:givesnv}
n(v_\phi,R)\,\d v_\phi=N(L_z)\,\d L_z\,P(R|L_z)
=N(L_z)P(R|L_z)R\d v_\phi,
\]
 where $N(L_z)\,\d L_z$ is the number of stars in the disc with $L_z$ in
 $(L_z,L_z+\d L_z)$.  In a cold disc, the number of stars with angular
momenta in $(L_z,L_z+\d L_z)$, is simply the mass in the corresponding
annulus, $2\pi\Sigma(\Rg)\Rg\,\d \Rg$.  Hence under the neglect of the mild
radial expansion noted above of a population as its dispersion increases,
an exponential disc with surface density
 \[\label{eq:expSig}
\Sigma(R)=\Sigma_0\e^{-(R-R_0)/R_\d},
\]
 where $\Sigma_0$ is the local surface density and $R_\d$ is the radial
scale-length of the disc, has
 \[\label{eq:givesN}
N(L_z)\simeq{2\pi\Sigma_0\Rg\over\vc}\,\e^{-(\Rg-R_0)/R_\d},
\]
 where we have assumed a constant circular speed, so $\d L_z=\vc\d\Rg$. 
Combining equations (\ref{eq:givesP}), (\ref{eq:givesnv}) and
(\ref{eq:givesN}), we have
\[\label{eq:nagain}
n(v_\phi,R)= {2\pi\Sigma_0\Rg\over\vc}\e^{-(\Rg-R_0)/R_\d}\,{K\over\sigma}
\exp\left[-{\Delta\Phieff(R,L_z)\over\sigma^2}\right].
\]

 Our assumption of constant $\vc$ allows us to  evaluate $K(\Rg)$, because
 then
\[
-\Delta\Phieff(R,L_z)=\fracj12L_z^2(\Rg^{-2}-R^{-2})+\vc^2\ln(\Rg/R),
\]
 so the normalisation condition $1=2\pi\int\d R\,RP$ reads
 \begin{eqnarray}\label{eq:normc}
{\sigma\over2\pi K}&=&\int\d
R\,\exp\left[-{\Delta\Phieff(R,L_z)\over\sigma^2}\right] \nonumber\\
&=&\int\d R\,\exp\left[c\left(2\ln\Rg/R+1-\Rg^2/R^2\right)\right]\\
&=&g(c)\Rg,\nonumber
\end{eqnarray}
 where
\[\label{eq:defsc}
c(\Rg)\equiv {\vc^2\over2\sigma^2(\Rg)},
\]
and 
\[\label{eq:defsg}
g(c)\equiv{\e^c(c-\frac32)!\over2c^{(c-1/2)}}.
\]
 So
\begin{eqnarray}\label{eq:noldvphi}
n(v_\phi,R)&=&{\Sigma_0\over\vc g(c)}
\exp\left[-{\Delta\Phieff\over\sigma^2}-{{\Rg-R_0}\over R_\d}\right]\nonumber\\
&=&{\Sigma_0\over\vc g(c)}
\exp\left[c\left(2\ln{\frac{\Rg}{R}}+1-\frac{\Rg^2}{R^2}\right)
-{{\Rg-R_0}\over R_\d}\right].
\end{eqnarray}
As here we are aiming at velocity distributions and not stellar densities at
a certain position, we will henceforth use the normalised velocity distribution at a fixed radius R
\begin{equation}\label{eq:nvphi}
n(v_\phi|R)= \frac{{\mathcal N}}{g(c)}\exp\left[c\left(2\ln{\frac{\Rg}{R}}+1-\frac{\Rg^2}{R^2}\right)
-{{\Rg-R_0}\over R_\d}\right] ,
\end{equation}
where ${\mathcal N}$ normalises the integral of $n$ in $v_\phi$ to unity.

Note that on the right side of eq.~(\ref{eq:nvphi}) the dependence on $v_\phi$ is carried by the
instances of $\Rg=Rv_\phi/\vc$ and by $c(\Rg)$.

\begin{figure}
\epsfig{file=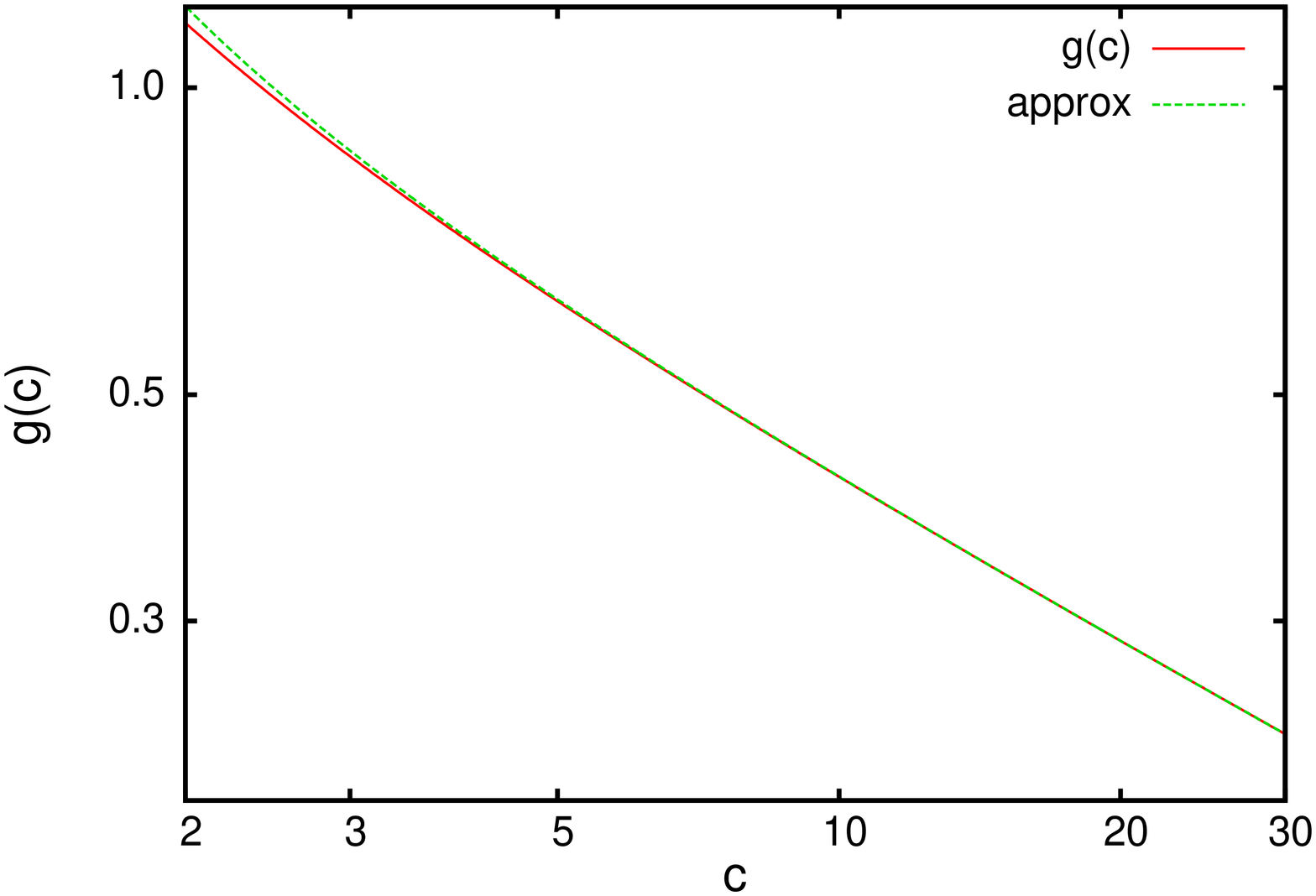,angle=-90,width=\hsize}
\epsfig{file=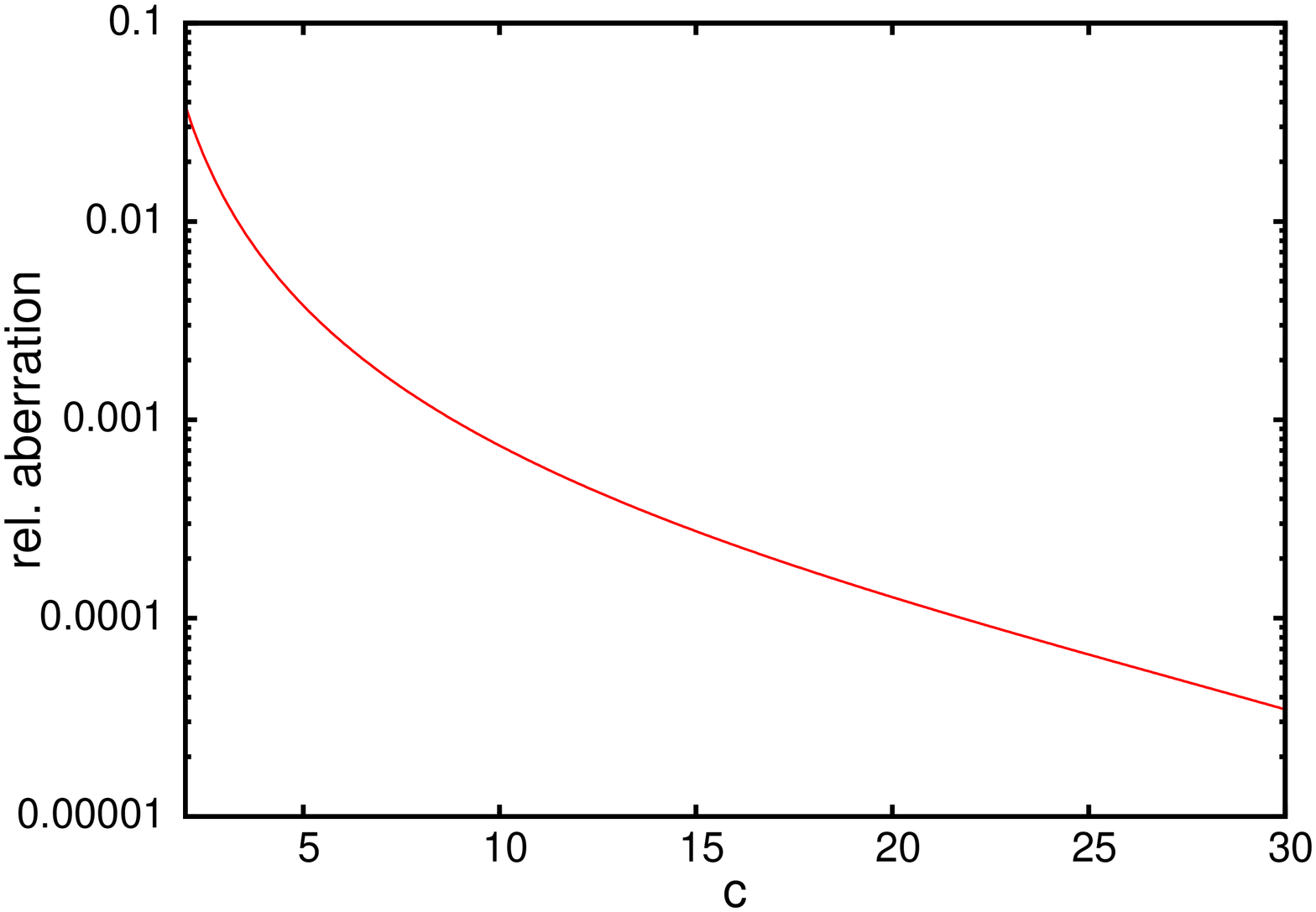,angle=-90,width=\hsize}
 \caption{Upper panel: The function $g(c)$ defined by eq.~(\ref{eq:defsg})
with solid red line and the approximation of eq.~(\ref{eq:appg}) shown
by a dashed green line. Lower panel:
the relative difference between the approximation and the underlying formula.}
\label{fig:prefac}
\end{figure}

From equation (\ref{eq:givesP}) we expect $K$ (which has the units of a
frequency) to be constant when $\sigma\ll\vc$, and indeed from an asymptotic
expansion of equation (\ref{eq:normc}) for large c we obtain $g(c)\propto
c^{-1/2}\propto \sigma$ and $K = \vc / (2\pi^{3/2} R_g)$.
\figref{fig:prefac} shows that $g(c)$ satisfies this expectation throughout
the entire parameter range of interest. For ($c > 25$), the direct
computation of $g(c)$ becomes impractical so apart from the advantage of a
numerically less costly formula a reasonable approximation must be found. The
dashed green line in \figref{fig:prefac} demonstrates that for $c>2$ this is
achieved to high precision by\footnote{Alternatively $g(c)$ can be
stringently approximated using eq.~(8.327) of \cite{GR}, but our term appears
to be the best compromise of simplicity and accuracy throughout the
interesting part of the parameter range.} 
 \[\label{eq:appg}
g(c)\simeq
\sqrt{\pi\over2(c-0.913)}.
\]
  For $\sigma$ we adopt the radial dependence
 \[\label{eq:sigdecline}
\sigma(\Rg)=\sigma_0\e^{-(\Rg-R_0)/R_\sigma}.
\]

Above we have restricted ourselves to stars with $E_z=0$. However, to the
extent that the motion in $R$ of a star is unaffected by its motion
perpendicular to the plane, the distribution we have derived will apply to the
population formed by all stars that now lie in the solar cylinder (the region
restricted in radius to $R\simeq R_0$ but unrestricted in $z$).  From our
formulae we have that the shape of this velocity distribution is controlled
by four parameters: the galactocentric radius of the measurement $R_0$, the
scale-length $R_\d$ of the young disc, the local velocity dispersion
$\sigma_0$, and the scale-length $R_\sigma$ on which the velocity dispersion
varies. The first two parameters are generally well-known and for fits of the
solar neighbourhood can be set to $R_0=8\kpc$ and $R_\d=2.5\kpc$. The value of
$R_\sigma$ is less clear and will be discussed below.

\begin{figure}
\centerline{\epsfig{file=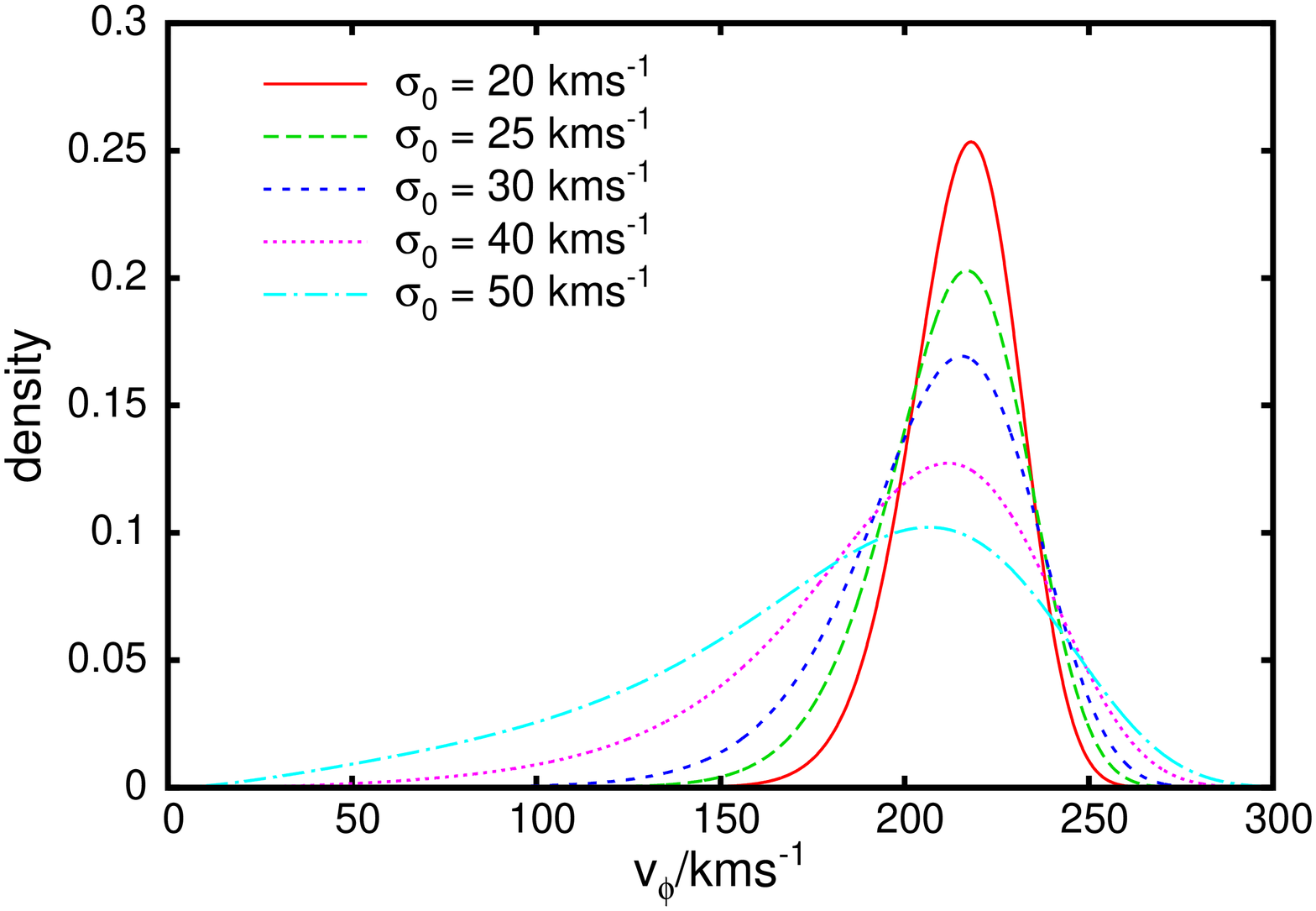,angle=-90,width=\hsize}}
\centerline{\epsfig{file=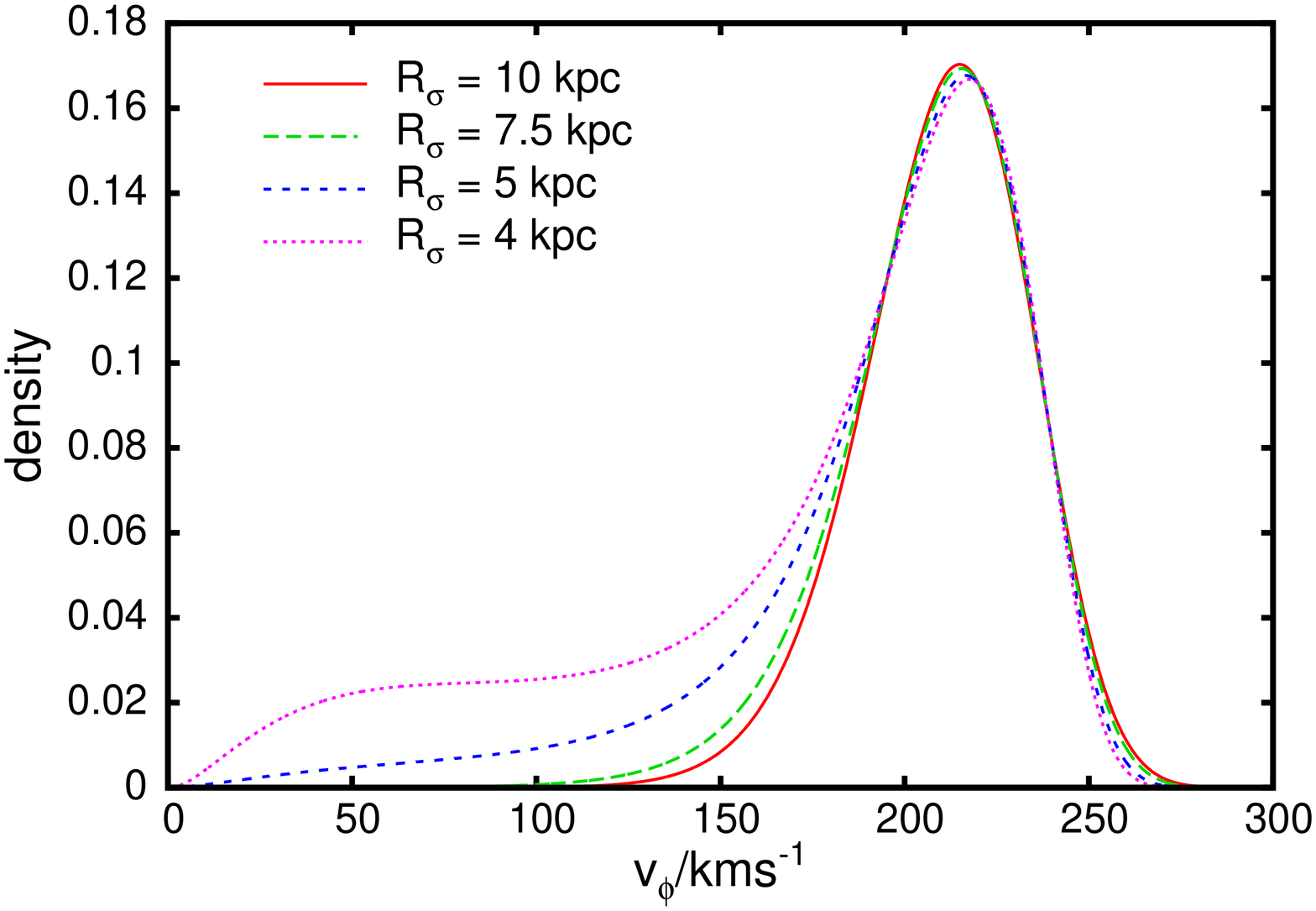,angle=-90,width=\hsize}} \caption{The
behaviour of eq.~(\ref{eq:nvphi}) for different local dispersions (upper
panel) and different scale-lengths on which the dispersion varies (lower
panel).  The radius
of observation is $R=8\kpc$ at a young-disc scale-length $R_\d = 2.5\kpc$. In
the upper panel we set the scale-length of the velocity dispersion to
$R_\sigma=3R_\d=7.5 \kpc$ and show results for local dispersions
$\sigma_0=20,25,30,40,50\kms$, while we use $\sigma_0=30\kms$ in the lower
panel.}\label{fig:var}
\end{figure}

The dependence of $n(v_\phi|R_0)$ on $\sigma_0$ and $R_{\sigma}$ is shown in
\figref{fig:var}.  In the upper panel we hold the dispersion scale-length
constant at $7.5\kpc$ and show the velocity distributions for local
dispersion values of $\sigma_0=20,25,30,40,50\kms$. As $\sigma_0$ increases,
the distribution becomes wider and the low-velocity tail rises much faster
than does the high-velocity tail, corresponding to an increasing asymmetric
drift.  Simultaneously, the peak slowly shifts to lower velocities. The lower
panel shows the velocity distributions for fixed $\sigma_0=30\kms$, but
different scale-lengths of the velocity dispersion,
$R_\sigma=10,7.5,5,4\kpc$.  Smaller values of $R_\sigma$ imply higher
dispersion in the inner regions of the Galaxy and lower dispersion in the
outskirts.  Thus for small $R_\sigma$ stars from the inner disc can more
easily reach the $R_0$ compared to their counterparts with larger $R_g$ so
the azimuthal velocity distribution is more skewed and develops a strong
low-velocity tail. In the most extreme case, $R_\sigma=4\kpc$, the model
breaks down, as $\ex{v_R^2}^{1/2}$ approaches $v_c$ in the inner regions.
Notice that as $R_\sigma$ falls, the mode of the velocity distribution shifts
to higher $v_\phi$ even as the tail at high $v_\phi$ becomes weaker. This
effect reflects the fact that stars with large random velocities spread their
contributions to the velocity distribution over many radii, while stars with
small random velocities localise their contributions. Consequently, when 
$R_\sigma$ is small, the width in $R$ over which stars contribute to $v_\phi$
narrows strongly as $\Rg$ increases, which puts stars with $\Rg\simeq R_0$
strongly in control of the mode of the local $v_\phi$ distribution.

\section{The velocity distribution as a function of distance from the plane}\label{sec:3d}

In the previous section we set $E_z=0$ to obtain results that are
approximately valid for the distribution of stars in $v_\phi$ regardless of
their distance from the plane. In practice we can determine the velocity
distributions of the stars that lie in various more-or-less narrow ranges in
$z$. We must now consider how these distributions will vary with $z$ and
differ from the aggregate distribution determined above. In the
following we do so using a number of quite crude physical approximations.
These  approximations give useful physical
insight into why the distribution in $v_\phi$ varies with $|z|$ as it does,
but ultimately the value of our final fitting formula does not depend on the
correctness of the arguments used to motivate it.

\subsection{Weights of different populations as functions of
$z$}\label{sec:getf}

The key idea is that within the solar cylinder there coexist many
populations, one for each value of $\Rg$. Indeed the chemical compositions
and ages of stars vary systematically with $\Rg$ \citep[see e.g.][for observations and a short discussion of consequences for studies of kinematics]{Luck11, Bensby11, SBD}. Moreover,
the smaller a population's value of $\Rg$, the larger will be its mean value of
$E_z$ and therefore the larger will be its vertical scale-height $h$
at radius $R$; here $h(\Rg,R)$ is the distance that at radius $R$ provides
the best fit to the vertical density profile of the population through
 \[\label{eq:sheight}
n_\Rg(R,z)\propto{1\over h}\e^{-|z|/h}.
\]
Note that with this formula we are not asserting that the real vertical
density profile is exponential, but simply identifying the characteristic
vertical extent of the population. We now investigate how the vertical
extent of the population increases with $R$ because the vertical restoring
force, which scales like $\Sigma(R)$, decreases outwards.  

BM11 show that a very good approximation to the vertical dynamics of a
population of stars can be obtained by assuming that the vertical action
\[
J_z\equiv\frac{1}{2\pi}\oint\d z\,v_z
\]
 of the population's stars is adiabatically invariant as the stars
oscillate in radius. We use this adiabatic approximation (\AA) to estimate the
ratio $h(\Rg,R)/h(\Rg,\Rg)$.

$J_z$ can be evaluated analytically only for a vertical force $K_z$ that is
proportional to $z^{\alpha-1}$ and to the local surface density of the disc
$\Sigma$, so we assume these dependencies in order to gain an analytic
model -- in reality $K_z$ has a much more complex dependence on $z$, which does
not yield an analytic expression for $E_z(J_z)$.
Then the vertical action is
 \[
J_z={2^{3/2}\over\pi}\int_0^{\zm}\d z\,\sqrt{E_z-k\Sigma z^\alpha},
\]
 where $k$ is a constant and 
 \[\label{eq:defszm}
\zm=(E_z/k\Sigma)^{1/\alpha}
\]
 is the
height at which the radical vanishes. In terms of the variable $\theta=z/\zm$
we have
 \[\label{eq:givesAA}
J_z={2^{3/2}\over\pi}(k\Sigma)^{1/2}\zm^{1+\alpha/2}
\int_0^1\d\theta\,\sqrt{1-\theta^\alpha}.
\]
 With the \AA, it now  follows that
\[
\zm\propto\Sigma^{-1/(2+\alpha)}.
\]
 The scale-height $h(\Rg,R)$ of the population will scale like $\zm$, so
 \[\label{eq:giveshratio}
{h(\Rg,R)\over
h(\Rg,\Rg)}=\left({\Sigma(R)\over\Sigma(\Rg)}\right)^{-1/(2+\alpha)}
=\exp\left(R-\Rg\over(2+\alpha)R_\d\right).
\]
 Photometry of edge-on spiral galaxies shows that the overall scale-height
varies very little with radius \citep{vanderKruit82}.
Since the velocity dispersion at radius $R$ will be
dominated by the population that has $\Rg=R$, we assume that 
\[\label{eq:defshzero}
h_0\equiv h(R,R)
\]
 is
independent of $R$. Hence
 \[\label{eq:giveshrr}
{h(\Rg,R)\over h(R,R)}=\exp\left({R-\Rg\over(2+\alpha)R_\d}\right).
\]

 For stars that make only small-amplitude vertical oscillations, $K_z\propto
z$ so $\alpha=2$. If the amplitude of a star's oscillations significantly
exceeds the local disc scale-height (but is none the less small compared to
$R$), a better approximation is that the disc is razor thin, so
$K_z\simeq\hbox{constant}$ and $\alpha=1$. For amplitudes not small
compared to the disc's scale-length, a yet smaller value of $\alpha$ is
appropriate. In the fits described below we assumed that $\alpha$ decreases
with $z$ according to
 \[\label{eq:givesalpha}
\alpha(z)=\cases{2-1.5 z/1.5\kpc&for $z\le1.5\kpc$\cr
0.5&otherwise.}
\]
These choices are educated guesses that produce useful results.

\begin{figure}
\epsfig{file=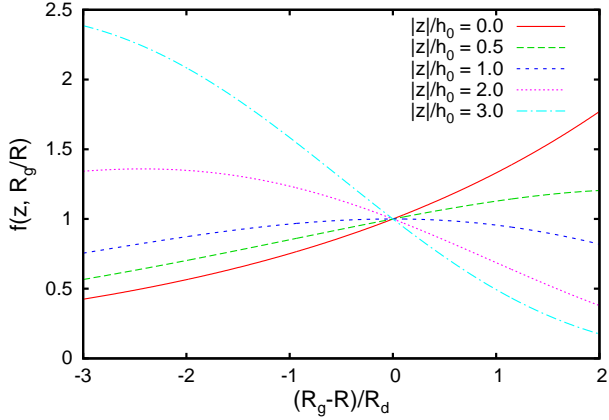,angle=-90,width=\hsize}
 \caption{The relative contributions given by eq.~(\ref{eq:fRg}) of the
populations with guiding-centre radius $\Rg$ measured at galactocentric radius $R$ to the velocity
distribution at various values of $|z|$. We use $\alpha = 1.5$, $h_0 = 300
\pc$ and $R_d = 2.5 \kpc$. 
}\label{fig:pref}
\end{figure}

Let the ratio of the contributions to the distribution in $v_\phi$ at height
$z$ of the population with guiding-centre radius $\Rg$ and the local
population be given by the factor $f(z,\Rg-R)$. Then from equation
(\ref{eq:noldvphi}) the distribution of $v_\phi$ at altitude $z$ above the plane is
 \begin{eqnarray}\label{eq:nvRzsimple}
n(v_\phi|R,z)&=&
\frac{{\mathcal N}}{g(c)}\exp\left[-(\Rg-R_0)\over R_\d\right]
\exp\left[-{\Delta\Phieff\over\sigma^2}\right]f(z,\Rg-R),
\end{eqnarray}
where ${\mathcal N}$ is a new normalisation constant.

By equation (\ref{eq:sheight}), the factor $f$ will scale as 
 \begin{eqnarray}\label{eq:fRg}
f(z,\Rg-R)&=&{n_\Rg(z)\over n_R(z)}\nonumber\\
&=&{h(R,R)\over h(\Rg,R)}\exp\left[{|z|\over h_0}\left(1-{h(R,R)\over
h(\Rg,R)}\right)\right]\\
&=&
\exp\left({\Rg-R\over (2+\alpha)R_\d}\right)\exp\left\{{|z|\over h_0}\left(1-
\exp\left({\Rg-R\over (2+\alpha)R_\d}\right)\right)\right\},\nonumber
\end{eqnarray}
 where in the second step we have used the definition (\ref{eq:defshzero}).
\figref{fig:pref} shows, for five values of $z$, how $f$ varies with $\Rg$ at
$R=R_0$. At $z=0$, $f$ is an increasing function of $\Rg$ because at large
$\Rg$ stars typically have small $E_z$ and therefore are more numerous in the
plane than in the solar cylinder as a whole. At $z=1\kpc$, by contrast, $f$
falls with increasing $\Rg$ because high above the plane a sample is richer
in stars with small $\Rg$, and therefore large $E_z$, than is the solar
cylinder as a whole. The formula attains a maximum, where the local
scale-height of a population equals the altitude. Since by definition
$f(z,\Rg-R_0)$ gives the enhancement of stars with guiding-centre radius
$\Rg$ relative to local stars, all lines intersect at ($f=1$, $\Rg=R_0$)
irrespective of the chosen altitude.

\subsection{Impact of the AA on the radial motion}\label{sec:backreact}

The \AA\ predicts that any star's value of $E_z$ varies as it moves radially.
Since the star is moving in a time-independent potential, its total energy is
constant. It follows that changes in $E_z$ must be compensated by changes in
its energy of motion parallel to the plane. Since $L_z$ is a true invariant,
the energy $\Phi_{\rm eff}(\Rg,L_z)$ associated with the underlying circular
orbit is unchanged, so changes in $E_z$ must be compensated by changes in the
radial energy $E_R$. Hence (\ref{eq:energy}) becomes
 \[
\fracj12v_R^2+\Delta\Phieff(R,L_z)+\Delta E_z=\hbox{constant},
\]
 where
\[\label{eq:defDEz}
\Delta E_z(J_z, R,\Rg)\equiv E_z(J_z,R)-E_z(J_z,\Rg):
\]
 takes into account the decrease in a star's vertical energy as it moves
outwards: by conservation of energy, this energy must be transferred to the
radial motion. One way of thinking about this energy transfer is to imagine
that the star moves at constant energy in an effective potential
 \[\label{eq:defPhiad}
\Delta\Phiad(R,L_z)\equiv\Delta\Phieff(R,L_z) + \Delta E_z(J_z, R,\Rg),
\]
 which might be called the ``adiabatic potential'' since it is the effective
potential for radial motion that follows from adiabatic invariance of the
vertical motion. At $R>\Rg$ $\Delta\Phiad$ increases with $R$ less rapidly
than the standard effective potential $\Delta\Phi_{\rm eff}$, so stars can reach
larger radii than they could if $E_z$ were constant.  BM11 simulated this
effect by simply increasing $L_z$ to $L_z+J_z$.

With the power-law forms of the vertical potential that we introduced above,
we can obtain the dependence of $E_z$ on $\Sigma$. From equations
(\ref{eq:defszm}) and (\ref{eq:givesAA}) we find
 \begin{equation}\label{eq:ezadi}
E_z \propto \Sigma^{2/2+\alpha},
\end{equation}
 so
\begin{equation}\label{eq:givesdez}
\Delta E_z(J_z,R,\Rg)
=E_z(J_z,R)\left[1-\exp\left({2(R-\Rg)\over(2+\alpha)R_\d}\right)\right].
\end{equation}
 Our plan is to use $\Delta E_z$ to modify our expression (\ref{eq:nagain})
for $n(v_\phi|R,z)$, which has no dependence on $v_z$ and therefore does not
specify a value of $J_z$ or $E_z$. Therefore in equation (\ref{eq:givesdez})
we now replace $E_z(J_z,R)$ by an estimate $\overline{E_z}$ of the typical
vertical energy of the stars that are encountered at $(R,z)$ with angular
momentum $L_z=Rv_\phi$. Equation (\ref{eq:Ezbar})
in the Appendix is an expression for $\overline{E_z}$ in terms of $h(\Rg,R)$ and the vertical
component of the gravitational potential.

\begin{figure}
\epsfig{file=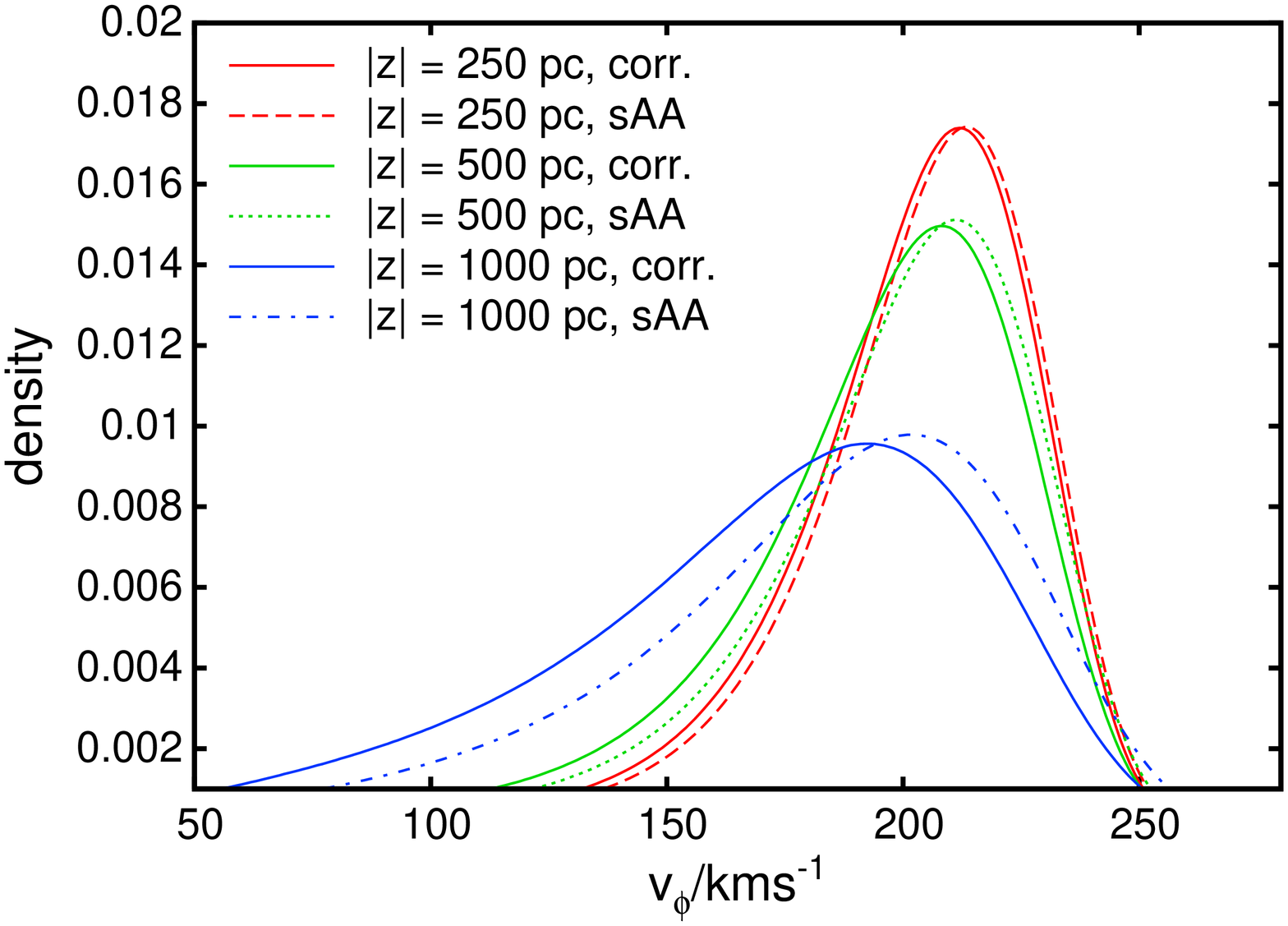,angle=-90,width=\hsize}
\epsfig{file=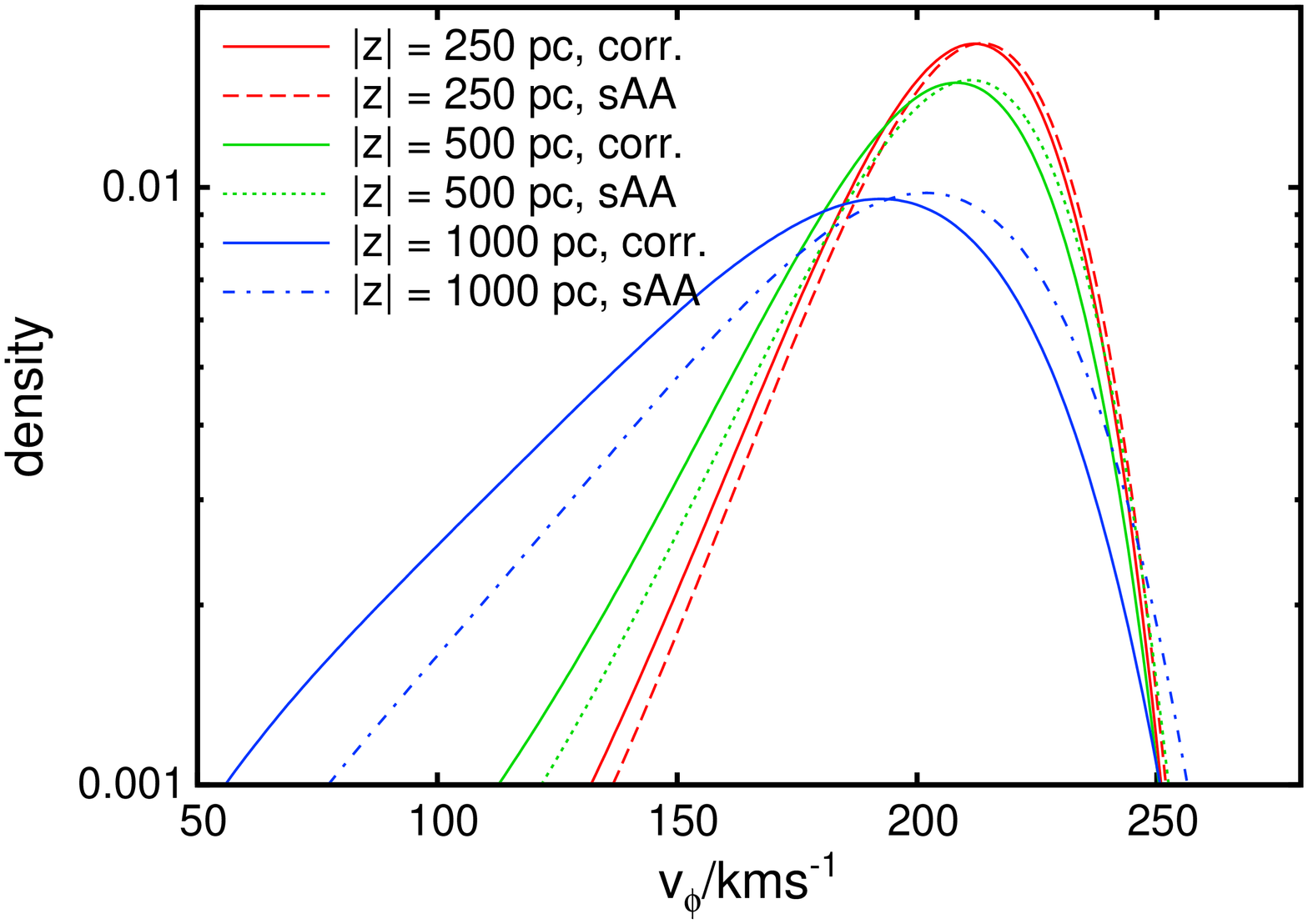,angle=-90,width=\hsize}
 \caption{Demonstration of the effect on velocity distributions above the Sun
of adding $\Delta E_z$ to the effective potential to allow for the tendency
of vertical energy to shift orbits outwards. The velocity distributions are
derived from fits to the models discussed in Section \ref{sec:apptorus}. Full
curves show the full corrections with $E_z$ (``corr"), dashed lines give the
same models with the simple adiabatic approximation (``sAA") ignoring energy
conservation.  }\label{fig:verten}
\end{figure}

Our next step is essentially to replace $\Delta\Phieff$ by $\Delta\Phiad$ in
equation (\ref{eq:nvRzsimple}) but before we do this we have to recall that
the prefactor $g(c)$ arose from normalising the radial probability density
$P(R|L_z)$ as given by equation (\ref{eq:givesP}). For consistency in this
formula we must now replace $\Delta\Phieff$ by $\Delta\Phiad$, with the
consequence that the normalising integral is no longer analytic. Therefore
our final formula for the distribution of $v_\phi$ at a given point in the
Galaxy must be written
 \begin{eqnarray}\label{eq:nvRz}
n(v_\phi|R,z)&=&
{\mathcal N}\e^{-(\Rg-R_0)/R_\d}\,{2\pi \Rg K\over\sigma}\nonumber\\
&&\times
\exp\left[-{\Delta\Phiad\over\sigma^2}\right]f(z,\Rg-R),
\end{eqnarray}
 where $K(c, E_z, R_g)$ is the numerically-determined result of normalising
the revised form of $P(R|L_z)$ [cf eq.~(\ref{eq:normc})], ${\mathcal N}$
normalises the distribution in $v_\phi$ and $f$ is defined by equation
(\ref{eq:fRg}).  

\figref{fig:verten} illustrates the effect that the
inclusion of $\Delta E_z$ in the effective potential has on $n(v_\phi|R,z)$
by showing for $R=R_0$ and several values of $z$ the velocity distributions
predicted with (full curves) and without (dotted curves) $\Delta E_z$.
Including $\Delta E_z$ moves all velocity distributions to lower $v_\phi$,
particularly on the left side.  The magnitude of the shift increases with $z$
because at low $z$, $n(v_\phi|R,z)$ is dominated by orbits with small $E_z$
and therefore small $\Delta E_z$.

Far from the plane ($z \sim 2 \kpc$) equation (\ref{eq:nvRz}) breaks down
because the physical assumptions on which it depends fail. A problem that
must be encountered at some height is that the
vertical frequency becomes comparable to the horizontal frequency, so the
assumption of adiabatic invariance of $J_z$ fails -- the coupling between the
horizontal and vertical motions becomes strong and complex. However, BM11
did not encounter problems with the \AA\ below $\sim z=2\kpc$. The failure we
encounter here probably arises from equation (\ref{eq:Ezbar}) for
$\overline{E_z}$, so we below explore the effect of
limiting the energy transfer by placing an uper limit on the value of
$\overline{E_z}$ to less than $(50\kms)^2$.

\subsection{Velocity moments}\label{sec:moments}

From the formulae we have in hand we can calculate a variety of moments.
Suppose the stellar population of the disc were a superposition of
populations that have \df s of the form 
 \[\label{eq:threedDF}
f(E_R,L_z,E_z)\propto\e^{-E_R/\sigma^2}\e^{-E_z/\sigma_z^2},
\]
 where the dependence on $E_z$ is inspired by equation (\ref{eq:sheight}).
 Then 
since this \df\ is a Gaussian in $v_R$, we would have $\langle
v_R^2\rangle=\sigma^2$. To each value of $\Rg$ and therefore $v_\phi$ we
could ascribe a fixed value for $\langle v_R^2\rangle=\sigma^2$, where in
general lower $v_\phi$ are connected to higher $\langle
v_R^2\rangle=\sigma^2$. Hence the
velocity dispersion of the entire disc could be obtained from the weighted
average
 \[\label{eq:givessr}
\left<{v_R^2}\right>(R,z) = 
\int\d v_\phi\,{n(v_\phi|R,z){\sigma}^2(Rv_\phi/\vc)}.
\]
 Similarly the asymmetric drift would be
\[
v_{\rm a}(R,z) 
= v_c - \int\d v_\phi\,{n(v_\phi|R,z)v_{\phi}}.
\]
 From the work of Section \ref{sec:backreact} we know that $E_R$ is not
strictly a constant of the motion on account of the transfer of energy
between the radial and vertical motions, so these formulae are only
approximate. We shall see that they are nevertheless useful.

 The \df\ (\ref{eq:threedDF}) is also a Gaussian in $v_z$ so naively we have
$\langle v_z^2\rangle=\sigma_z^2$ for the population formed by stars of a
given value of $L_z$. However, when calculating $\langle v_z^2\rangle$ for
the entire disc it is essential to bear in mind the radial variation of $E_z$
implied by adiabatic invariance of $J_z$ (eq.~\ref{eq:ezadi}). In effect this
variation of $E_z$ causes $\sigma_z$ to vary with radius even at fixed $L_z$.
Hence an approximate expression for the vertical velocity dispersion of the entire disc
is
 \[
\left<v_z^2\right>(R,z) =
\int\d v_\phi\,
{n(v_\phi|R,z){\sigma_z}^2(R_0v_{\phi}/\vc)}
\left(\frac{\Sigma_0}{\Sigma(R)}\right)^{{2/2+\alpha}}.
\]

\section{Applications}

\subsection{Comparison with torus models}\label{sec:apptorus}

To evaluate the performance of the above equations we fit the
velocity distributions of the torus model of BM11. The \df\ of this model is
\[\label{totalDF}
f(J_r,L_z,J_z)=f_{\sigma_r}(J_r,L_z)\times
{\nu
\over2\pi\sigma_z^2}\,\e^{-\nu J_z/\sigma_z^2},
\]
where
 \[\label{planeDF}
f_{\sigma_r}(J_r,L_z)\equiv{\Omega\Sigma\over\pi\sigma_r^2\kappa}\bigg|_{\Rc}
[1+\tanh(L_z/L_0)]\e^{-\kappa J_r/\sigma_r^2}.
\]
 Here $\Omega(L_z)$ is the circular frequency for angular momentum $L_z$,
$\kappa(L_z)$ is the radial epicycle frequency and $\nu(L_z)$ is its vertical
counterpart and $\Sigma(\Rg)$ is given by equation (\ref{eq:expSig}).
The factor $1+\tanh(L_z/L_0)$ in
equation (\ref{planeDF}) is there to effectively eliminate stars on
counter-rotating orbits and the value of $L_0$ is unimportant provided it is
small compared to the angular momentum of the Sun. In equations
(\ref{totalDF}) and (\ref{planeDF}) the functions $\sigma_z(L_z)$ and
$\sigma_r(L_z)$ control the vertical and radial velocity dispersions. The
observed insensitivity to radius of the scale-heights of extragalactic discs
motivates the choices
 \begin{eqnarray}\label{eq:sigmas}
\sigma_r(L_z)&=&\sigma_{r0}\,\e^{q(R_0-\Rc)/R_\d}\nonumber\\
\sigma_z(L_z)&=&\sigma_{z0}\,\e^{q(R_0-\Rc)/R_\d},
\end{eqnarray}
 where $q=0.45$ and $\sigma_{r0}$ and $\sigma_{z0}$ are approximately equal
to the radial and vertical velocity dispersions at the Sun. BM11 take the \df\
of the entire disc to be the sum of a \df\ of the form (\ref{totalDF}) for
the thin disc, and a similar \df\ for the thick disc, the normalisations
being chosen so that at the Sun the surface density of thick-disc stars is
23 per cent of the total stellar surface density. Table \ref{tab:df} lists the
parameters of each component of the \df.

\begin{table}
\caption{Parameters of the torus model's \df.}\label{tab:df}
\begin{center}
\begin{tabular}{l|cccc}
Disc & $R_\d/\hbox{kpc}$ & $\sigma_{r0}/\!\kms$ & $\sigma_{z0}/\!\kms$ &
$L_0/\!\kpc\kms$\\
\hline
Thin & 2.4 & 27 & 20 & 10\\
Thick& 2.5 & 48 & 44 & 10\\
\end{tabular}
\end{center}
\end{table}

\begin{table}
\caption{Parameters of the potential employed}\label{tab:pot}
\begin{center}
\begin{tabular}{l|cccccc}
\hline
Component & $\Sigma(R_0)/\msun \pc^{-2}$ & $R_\d/\hbox{kpc}$ & $h/\!\kpc$ & $R_{\rm m}/\!\kpc$\\
\hline
Thin &36.42& 2.4 & 0.36 & 0\\
Thick&4.05& 2.4 & 1 & 0\\
Gas &8.36& 4.8 & 0.04 & 4\\
\hline
\end{tabular}
\begin{tabular}{l|cccccc}
Component & $\rho/\!\msun\pc^{-3}$ & $q$ & $\gamma$ & $\beta$ & $r_0/\!\kpc$
& $r_t/\!\kpc$\\
\hline
Bulge&0.7561&0.6&1.8&1.8&1&1.9\\
Halo&1.263&  0.8&$-2$&2.207&1.09&1000\\
\end{tabular}
\end{center}
\end{table}

\begin{table}
\begin{tabular}{c|cccccc}
$z$&$\sigma_0$&$R_{\sigma}$&$h_0$&$\langle v_R^2\rangle^{1/2}$&
$\langle
v_R^2\rangle^{1/2}_{\rm BM11}$&$\chi^2$\\ \hline
$0$ & $27.07$ & $5.30$ & --- &$30.9$& $33.4$& $0.000095$\\
$250$ & $27.56$ & $5.39$&$229.8$ & $32.3$ & $35.2$ & $0.000134$\\
$500$ & $29.23$ & $5.70$& $170.4$& $36.4$ & $40.4$& $0.000265$\\
$750$ & $33.30$ & $6.25$ & $190.4$ & $44.1$ & $48.4$ & $0.000338$\\
$1000$ & $38.93$ & $6.86$ & $247.7$ & $53.9$ & $55.1$&  $0.000063$\\
$1250$ & $40.85$ & $6.90$ & $281.3$ & $60.0$ & $58.7$ & $0.000079$\\
$1500$ & $42.21$ & $7.29$ & $294.5$ & $62.4$ & $60.7$ & $0.000034$ \\
$1750$ & $44.45$ & $8.26$ & $284.7$ & $63.7$ & $61.8$& $0.000113$ \\
$2000$ & $49.27$ & $10.90$ & $271.5$ & $64.8$ & $63.1$ & $0.000363$ \\
\end{tabular}
 \caption{Values of the parameters for the fits shown in
\figref{fig:fitBM10old}, which employ the simple adiabatic approximation. $z$ denotes the distance from the plane in parsecs,
$R_{\sigma}$ the scale-length of radial velocity dispersion, $h_0$ the
local scale-height. We fixed the circular speed $v_c = 216.25 \kms$ and disc
scale-length $R_\d = 2.4 \kpc$ to the values used by BM11. For a further
comparison we give the rms radial velocity from eq.~(\ref{eq:givessr}) and
the value of the corresponding parameter $\sigma_{R,\rm BM11}$ of the torus
model. All fits and their $\chi^2$ values were derived for $60 \kms < v_\phi < 260 \kms$.
Note that the $\chi^2$ values are not sensible per se, but a mere description
of the relative fit quality, as there are no proper errors on the theoretical
distributions underlying the fit.}\label{tab:bpmold}
\end{table}

\begin{figure}
\epsfig{file=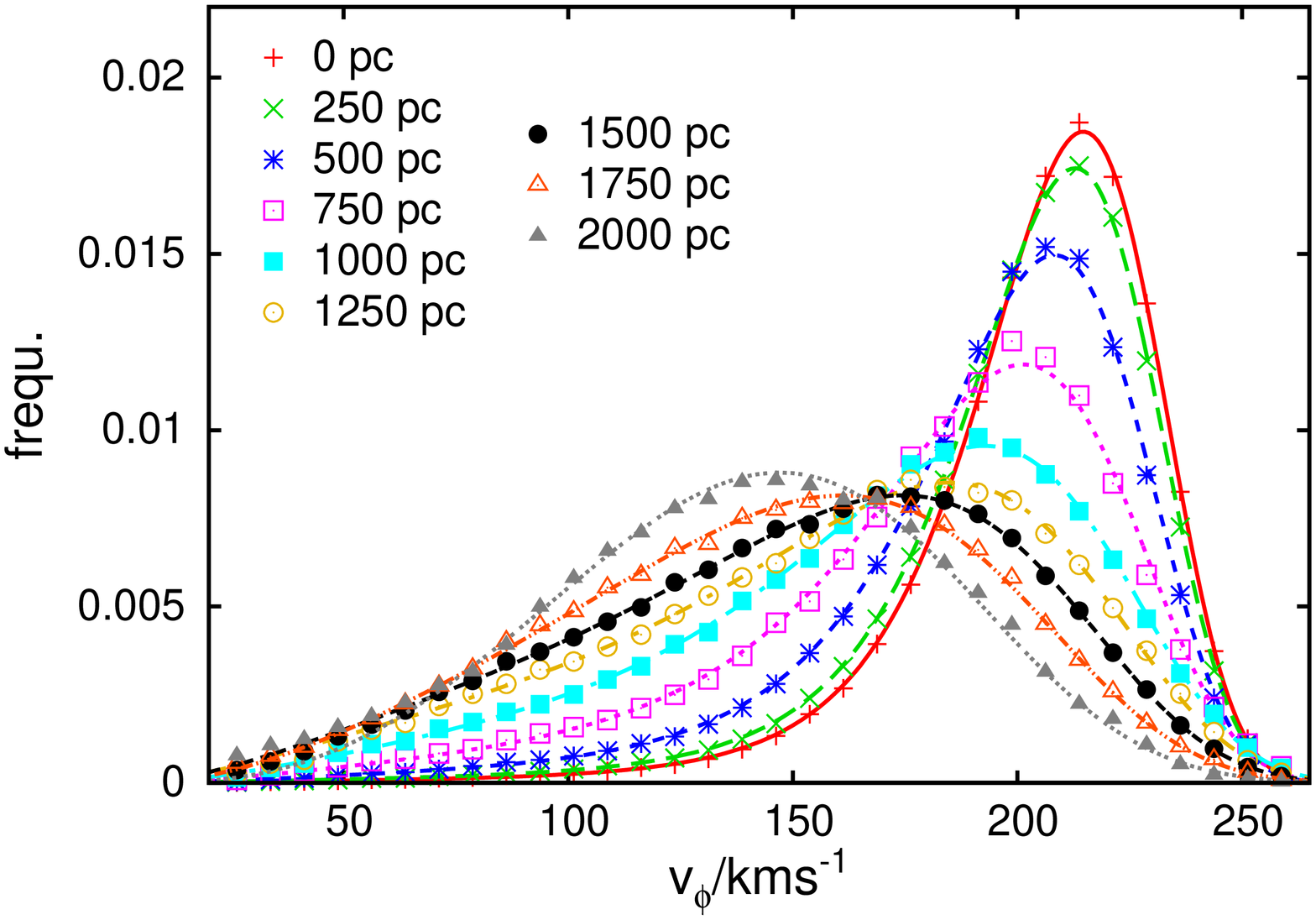,angle=-90,width=\hsize}
\epsfig{file=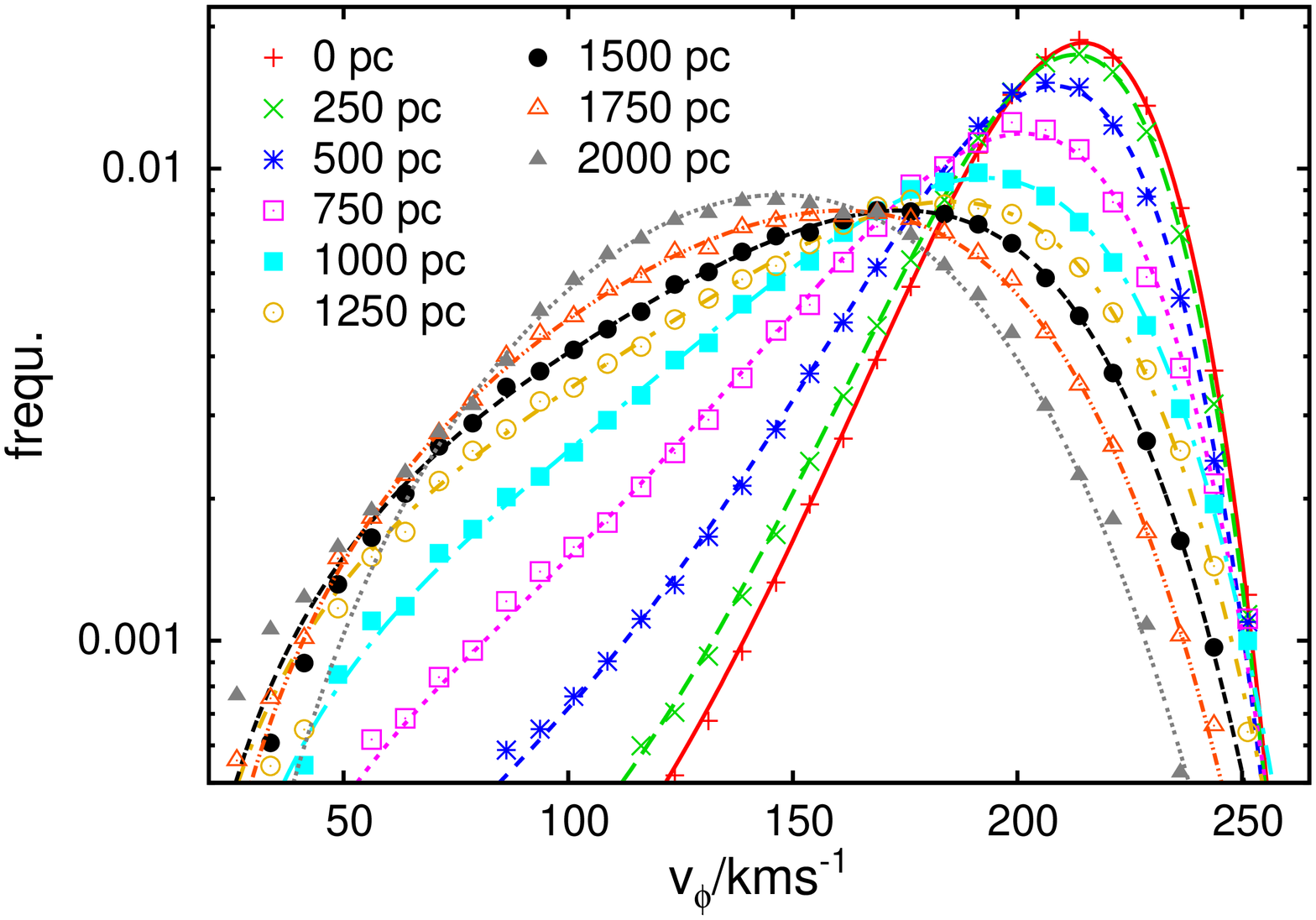,angle=-90,width=\hsize}
 \caption{Fitting velocity distributions of a torus model from BM11 at
different altitudes using eq.~(\ref{eq:nvRzsimple}). Points show values from
the torus model and the curves show our fits to these points. The
scale-length of the disc was fixed at the BM11 value of $R_\d = 2.4 \kpc$, the
parameter $\alpha$ used to estimate the adiabatic invariant was set via equation (\ref{eq:givesalpha}).}\label{fig:fitBM10old}
\end{figure}

BM11 took the gravitational potential to be that of Model 2 in
\cite{DehnenB97} modified to have thin- and thick-disc scale-heights of
$360\pc$ and $1\kpc$ (Table~\ref{tab:pot}). In this model the disc
contributes 60 per cent of the gravitational force on the Sun, with dark
matter contributing most of the remaining force.

\subsubsection{Distribution in $v_\phi$}

\begin{figure}
\epsfig{file=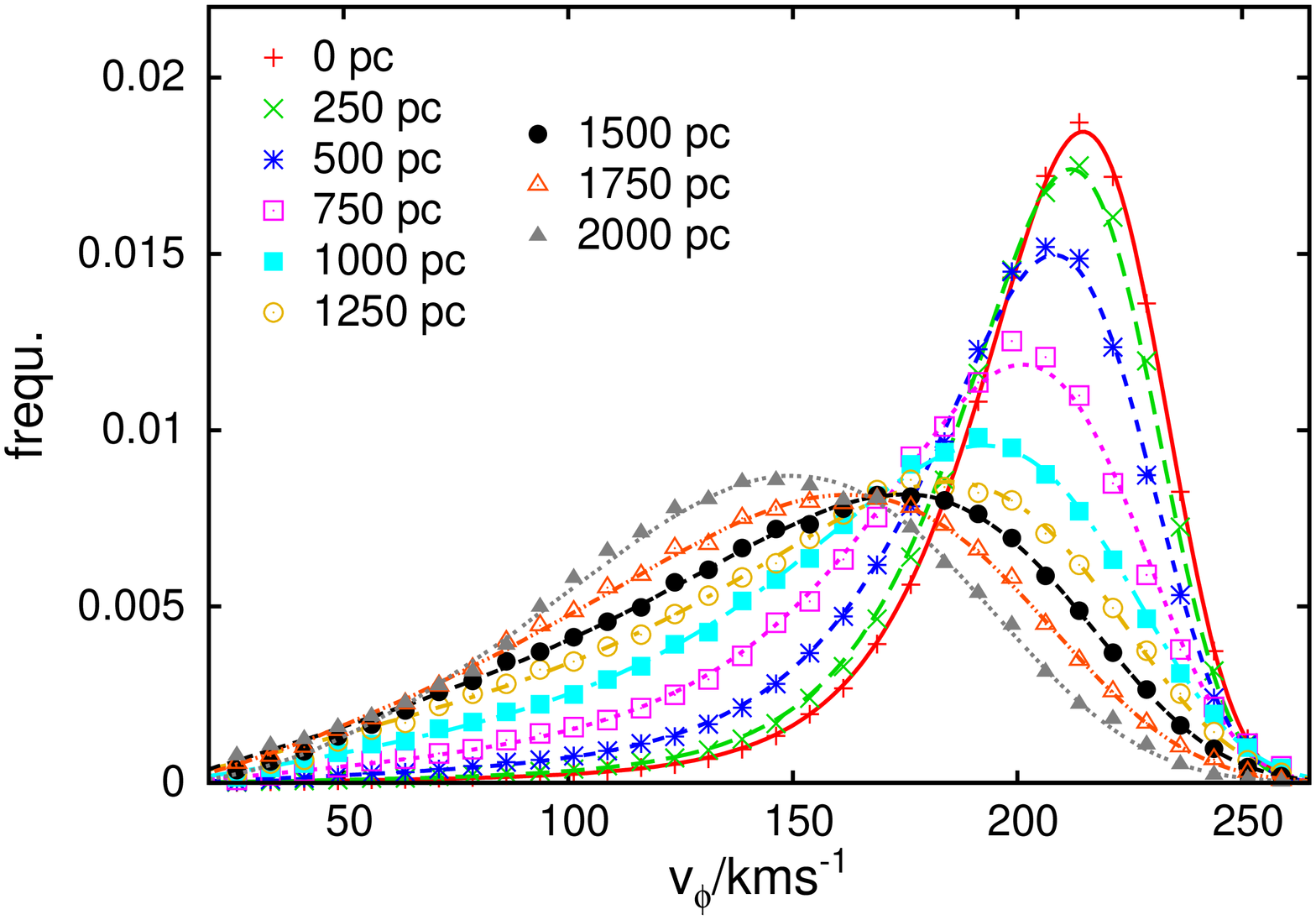,angle=-90,width=\hsize}
\epsfig{file=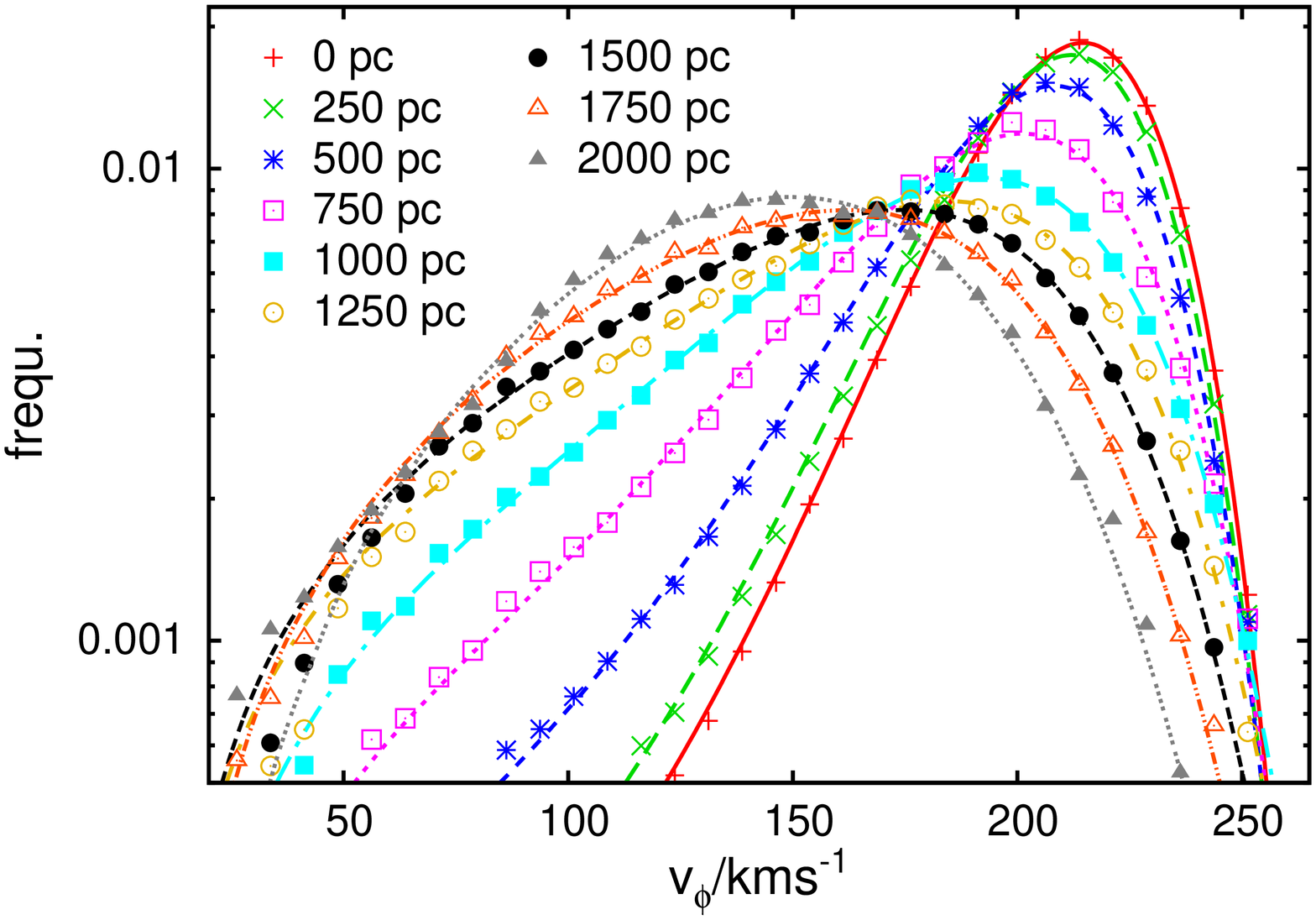,angle=-90,width=\hsize}
 \caption{Fitting the velocity distributions of BM11 at
different altitudes using eq.~(\ref{eq:nvRz}). The scale-length of the disc was fixed at their value of
$R_\d = 2.4 \kpc$ and the parameter $\alpha$ was assumed to be the function of
$z$ specified by eq.~(\ref{eq:givesalpha}).}\label{fig:bpezf}
\end{figure}

The points in \figref{fig:fitBM10old} show the $v_\phi$ distributions of the
BM11 model at $R_0$ and heights up to $2\kpc$, while the curves show the fits
to these points that we obtain from equation (\ref{eq:nvRz}) when we
approximate $\Delta\Phiad$ with $\Delta\Phieff$. We take $\alpha$ from
equation (\ref{eq:givesalpha}) and at each location $(R_0,z)$ we fit the
given $v_\phi$ distribution independently by adjusting the parameters
$\sigma_0$, $R_\sigma$ and $h_0$.  Table~\ref{tab:bpmold} gives the parameter
values obtained from the fits and also the radial velocity dispersions that
the fits yield through equation (\ref{eq:givessr}) and the true radial
velocity dispersion within the model. Notice that the parameter $\sigma_0$
rises from $27\kms$ at the plane to $48\kms$ at $z\sim2\kpc$ and that these
values coincide with the values of the corresponding parameters in the \df s
of the thin and thick discs, respectively.  \figref{fig:bpezf} and Table
\ref{tab:bpez} show the fits obtained to the same data when $\Delta\Phiad$ in
equation (\ref{eq:nvRz}) is evaluated from equation (\ref{eq:defPhiad}) with
$\Delta E_z$ replaced by $\overline{E_z}$ from equation (\ref{eq:Ezbar}).

In both Figs.~\ref{fig:fitBM10old} and \ref{fig:bpezf} the quality of the
fits is excellent, so the inclusion of $\Delta E_z$ improves the optimum fit
only marginally. However, inclusion of $\Delta E_z$ does change the optimum
value of $h_0$ significantly and in the sense of bringing it closer to the
true scale-height of the model disc, which increases from small values very
close to the plane, where the gas disc dominates the gravitational potential,
through $300\pc$ at $z\sim300\pc$, where the thin disc accounts for the
majority of stars, to $\sim1\kpc$ at large heights, where the thick disc is
dominant. Note that far from the plane the dominant population's scale-height
$h_0$ is significantly smaller than the locally measured scale-height of the
disc because the population is dominated by stars with $\Rg< R$, which by
equation (\ref{eq:giveshratio}) have $h(\Rg,R)> h_0$.  Even so, when
$\Delta E_z$ is omitted, the fitted values of $h_0$ are unexpectedly small.
Including $\Delta E_z$ increases $h_0$ at all heights, while limiting the
values of $\overline{E_z}$ employed to less than $(50\kms)^2$ yields
intermediate values of $h_0$, which are not far from constant as we would
wish. The reason adding $\Delta E_z$ to the effective potential increases
$h_0$ is that $\Delta E_z$ increases the contribution to the velocity
distribution at $R_0$ of stars with small values of $\Rg$ and therefore
$v_\phi$ and thus reduces the need to suppress the contribution of the
population with $\Rg\simeq R_0$, which dominates the peak of the $v_\phi$
distribution, relative to the stars that form the prominent left wing of the
distribution. The other improvement effected by including $\Delta E_z$ is to
lower $\langle v_R^2\rangle^{1/2}$ slightly and thus bring it closer to its
true value at high altitudes. When there is no upper limit on the values of
$\overline{E_z}$ used in the calculation of $\Delta E_z$,  this lowering of
$\langle v_R^2\rangle^{1/2}$ becomes excessive above $z\simeq2\kpc$ because
at such  altitudes the vertical energy becomes comparable to the radial
energy.

Irrespective of whether $\Delta E_z$ is used, the scale-length $R_\sigma$
exhibits a continuous rise in the fits, moving away from the value
$R_\d/0.45$ of the corresponding parameter of the torus model.  $R_{\sigma}$
comes closer to $R_\d/0.45$ the lower $\alpha$ is chosen at higher
altitudes.

\begin{table}
\begin{tabular}{c|cccccc}
$z$&$\sigma_0$&$R_{\sigma}$&$h_0$&$\langle v_R^2\rangle^{1/2}$&$\langle
v_R^2\rangle^{1/2}_{\rm BM11}$&$\chi^2$\\ \hline
$0$ & $27.06$ & $5.31$ &$50.5$&$30.9$& $33.4$& $0.000097$\\
$250$ & $27.46$ & $5.48$&$347.2$ & $32.1$ & $35.2$ & $0.000310$\\
$500$ & $29.65$ & $5.61$& $323.4$& $37.1$ & $40.4$& $0.000267$\\
$750$ & $33.99$ & $6.14$ & $326.8$ & $45.4$ & $48.4$ & $0.000338$\\
$1000$ & $39.81$ & $6.79$ & $438.7$ & $55.4$ & $55.1$&  $0.000060$\\
$1250$ & $41.99$ & $6.81$ & $533.2$ & $62.3$ & $58.7$ & $0.000085$\\
$1500$ & $43.08$ & $7.40$ & $564.1$ & $63.6$ & $60.7$ & $0.000040$ \\
$1750$ & $43.83$ & $8.50$ & $496.4$ & $62.3$ & $61.8$& $0.000117$ \\
$2000$ & $43.94$ & $10.70$ & $416.2$ & $58.3$ & $63.1$ & $0.000234$ \\
\hline
$0$ & $27.06$ & $5.31$ &$50.5$&$30.9$& $33.4$& $0.000097$\\
$250$ & $27.46$ & $5.48$&$347.2$ & $32.1$ & $35.2$ & $0.000310$\\
$500$ & $29.65$ & $5.61$& $323.4$& $37.1$ & $40.4$& $0.000269$\\
$750$ & $33.99$ & $6.14$ & $326.8$ & $45.4$ & $48.4$ & $0.000338$\\
$1000$ & $39.88$ & $6.74$ & $391.0$ & $55.7$ & $55.1$&  $0.000062$\\
$1250$ & $41.74$ & $6.79$ & $401.8$ & $61.9$ & $58.7$ & $0.000081$\\
$1500$ & $42.72$ & $7.28$ & $390.7$ & $63.4$ & $60.7$ & $0.000035$ \\
$1750$ & $43.69$ & $8.13$ & $345.8$ & $63.1$ & $61.8$&$0.000105$ \\
$2000$ & $45.87$ & $10.18$ & $301.9$ & $61.7$ & $63.1$ & $0.000230$ \\

\end{tabular}
 \caption{Fit parameters when the vertical energy correction $\Delta E_z$ is
included. The upper half applies the unlimited correction and its fits are
presented in \figref{fig:bpezf}. The lower half includes an upper limit
$\Delta E_z\simeq\overline{E_z} =(50\kms)^2$ that produces comparable fits,
but prevents a breakdown of the horizontal dispersion starting around $z = 2
\kpc$. Fits were taken with the same fixed parameters and in identical range
as the fits described in Table \ref{tab:bpmold}, but this time with non-zero
$\Delta E_z$.} \label{tab:bpez}
\end{table}

\begin{figure}
\epsfig{file=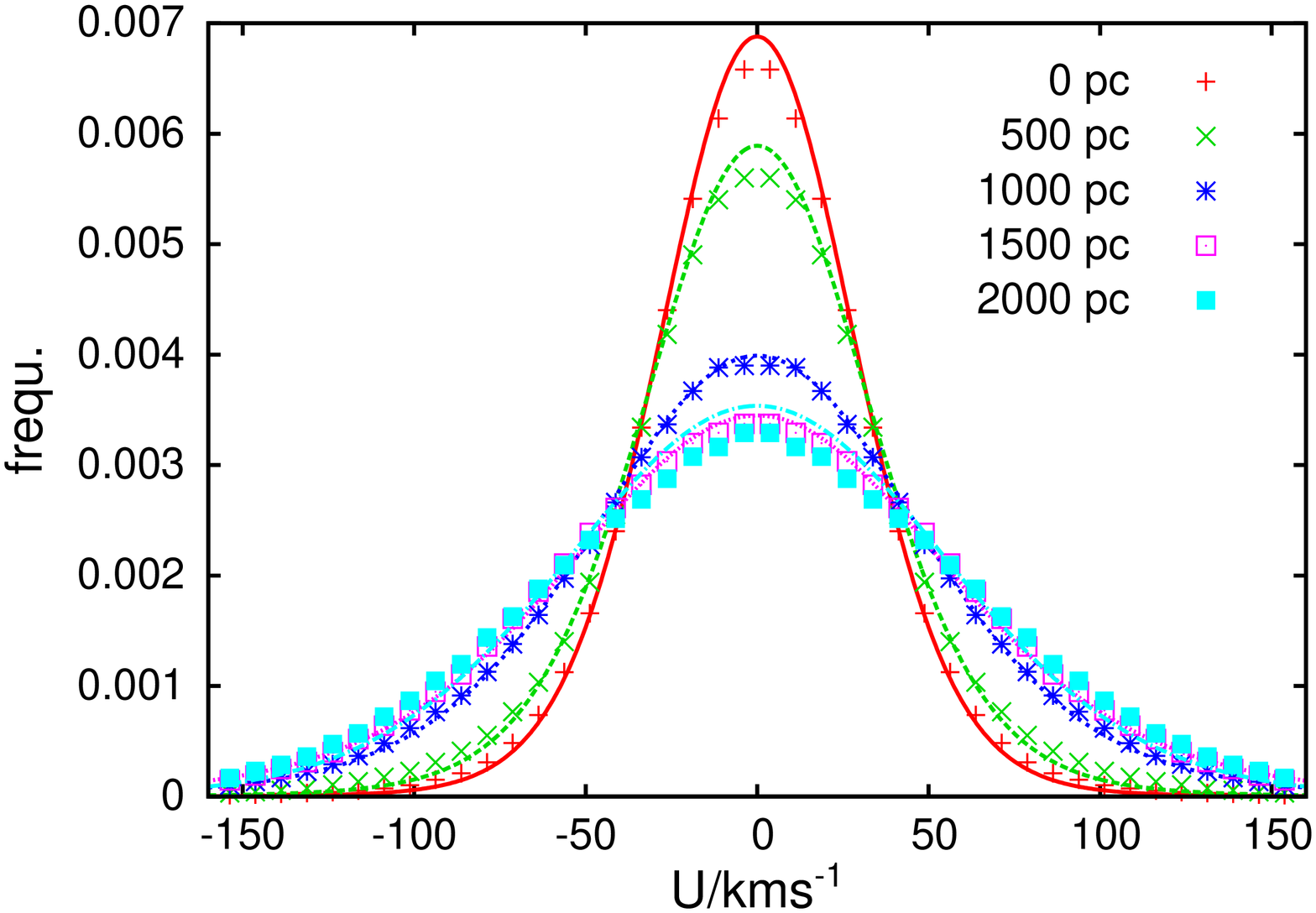,angle=-90,width=\hsize}
\epsfig{file=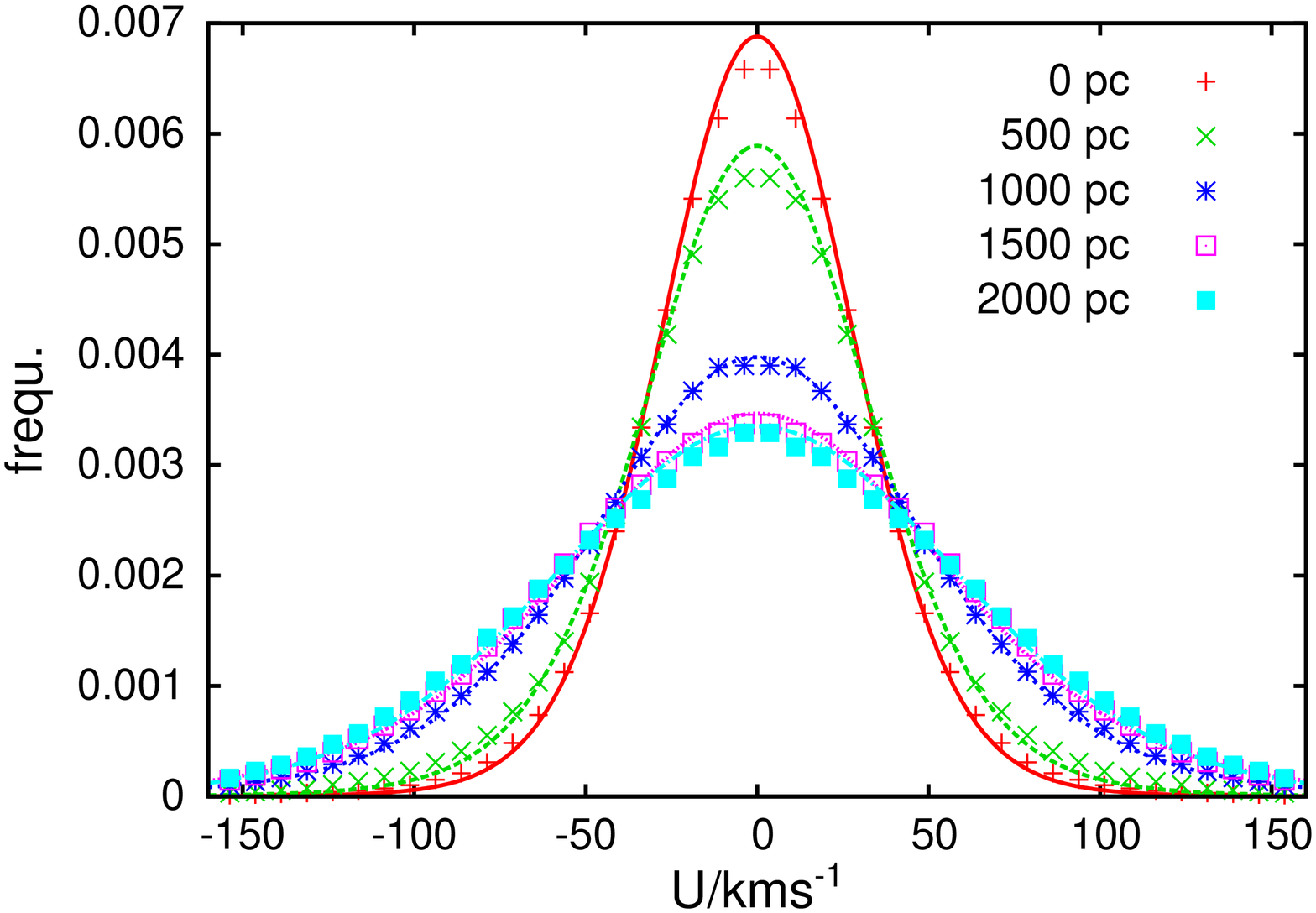,angle=-90,width=\hsize}
 \caption{$v_R$ velocity distributions from the models with non-zero $\Delta
E_z$ (lines) compared to the torus models (data points) at different
altitudes $z$. The upper panel is for when $\overline{E_z}$ is unlimited and
the lower panel is for when $\overline{E_z}<(50\kms)^2$.}\label{fig:Udistr}
\end{figure}

\subsubsection{Distributions in $v_R$}\label{sec:VR}

As we remarked in Section \ref{sec:moments}, the \df\ (\ref{eq:threedDF}) is
such that stars of given $L_z$ have a Gaussian distribution in $v_R$.
Consequently, the distribution in $v_R$ of all stars found at a given
distance from the plane should in this picture be a weighted sum of Gaussian
distributions with the weights implicit in equation (\ref{eq:givessr}).
Fig.~\ref{fig:Udistr} compares this prediction (lines) at several altitudes
$z$ with the corresponding distributions from the torus models (data points).
The overall agreement between the data points and the predictions of the
formula is remarkable when one bears in mind that the curves have not been
obtained by fitting to the data points.  At low altitudes (red and green) the
formula predicts a distribution that is slightly too sharply peaked and
deficient in the wings. Around $\sim1\kpc$ from the plane the fit is near
perfect. The agreement between the curve for $2\kpc$ and the data points at
$|v_R|<50\kms$ is much better in the lower panel than the upper panel,
vindicating  the use of the correction provided by $\Delta E_z$.

\begin{figure}
\epsfig{file=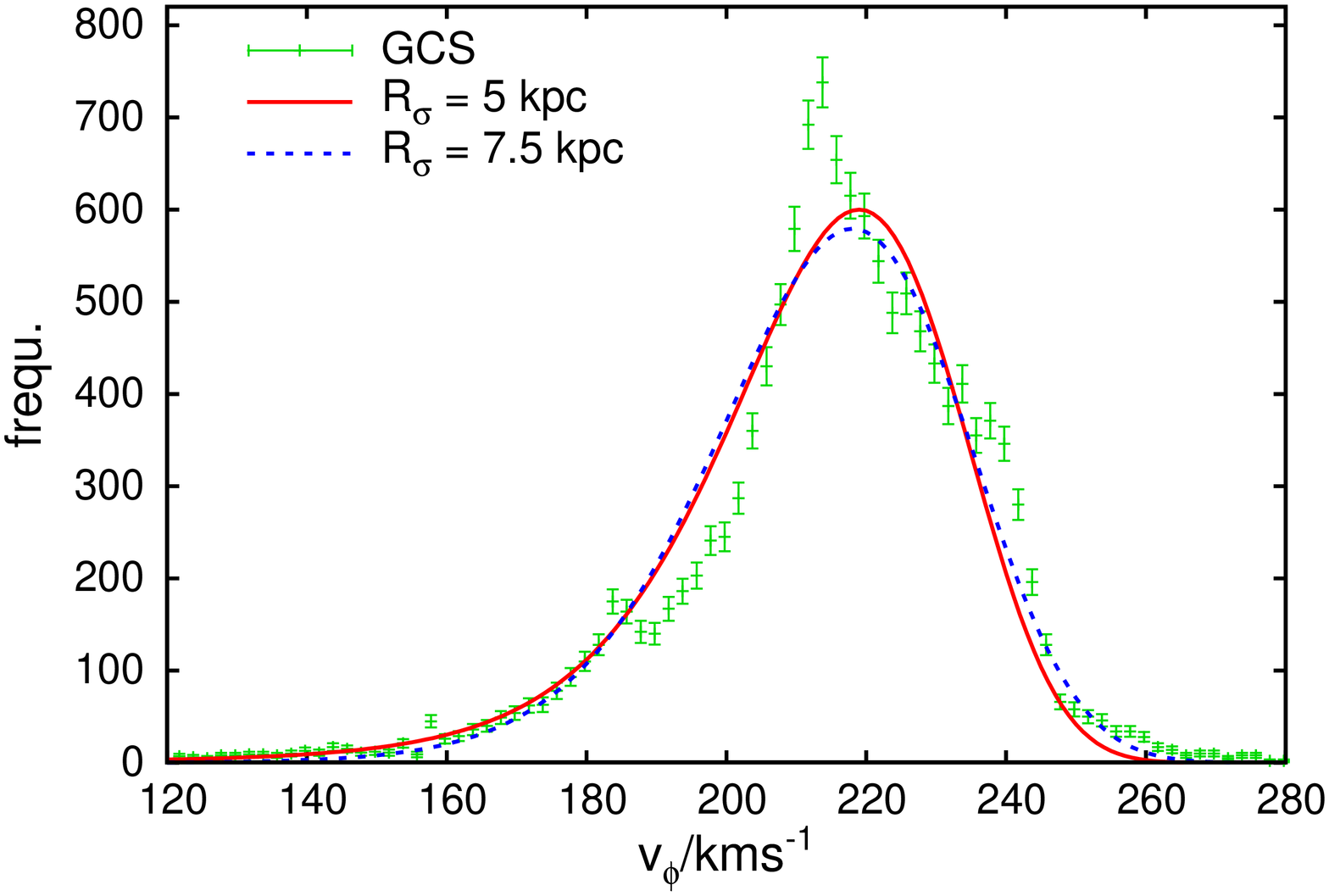,angle=-90,width=\hsize}
\epsfig{file=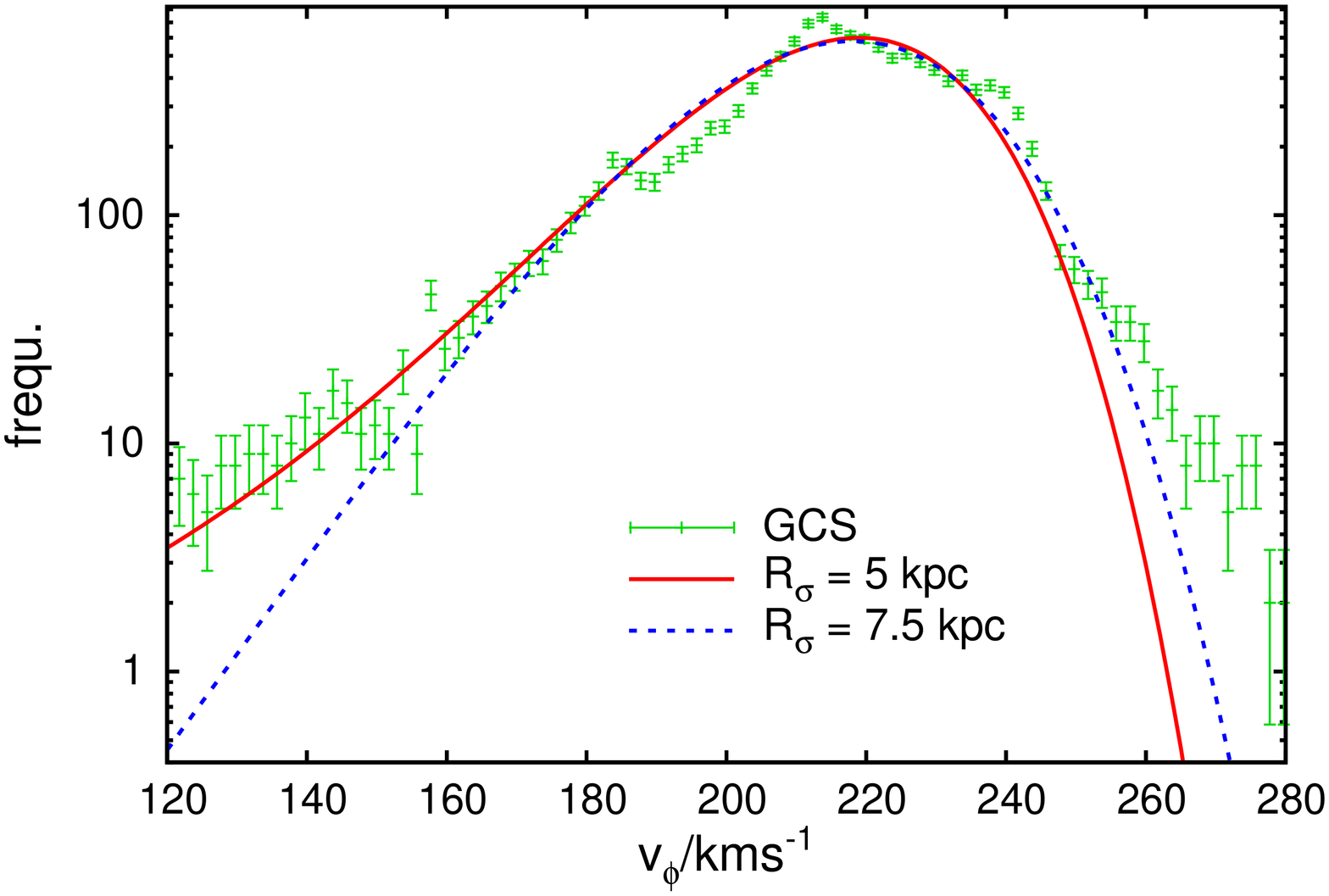,angle=-90,width=\hsize}
\caption{Fitting the GCS velocity distribution with a two-parameter
  fit, using only a normalization constant and the local dispersion
  $\sigma_0$. The lower panel shows the fit on a logarithmic scale to
  show the wings, while the upper panel presents the linear scale. We
  separately fit $\sigma_0$ for dispersion scale-length $R_{\sigma}=7.5\kpc$
(blue dashed line) and $R_{\sigma}=5\kpc$ (red solid line). 
}\label{fig:GCSfit}
\end{figure}

\subsection{Application to the Geneva-Copenhagen Survey}\label{sec:GCS}

We have fitted the full formula (\ref{eq:nvRz}) at an altitude of $z=40 \pc$
to the velocity distribution of the Geneva-Copenhagen Survey (GCS) of F and G
stars \citep[][]{Nord04, Holmberg09}. As sample we selected the full
$13\,520$ objects that have measured space velocities. We adopted
$\vc=220\kms$ and assumed that the Sun's velocity with respect to the Local
Standard of Rest is $12.24\kms$ \citep{SBD}, so $232.24\kms$ was added to the
published heliocentric $v_\phi$ velocities.  Given that the sample lies near
to the mid-plane, where $\alpha=2$ would apply, we adopted $\alpha = 1.5$,
and we also set the local scale-height to $h_0 = 200 \pc$: when $\alpha$ and
$h_0$ are allowed to vary when fitting to the data, they prove to be strongly
correlated in the sense that low values of $\alpha$ enhance the contribution
of populations with smaller $\Rg$ and thus favour larger values of $h_0$ for
balance. However, the differences between the residuals of the various best
fits are not statistically significant.

\figref{fig:GCSfit} shows two typical fits performed on the region $150 <
v_\phi/\kms < 250$, which demonstrate how nicely and naturally the formula
reproduces the non-Gaussianity of the azimuthal velocity distribution.  The
fits are for scale-lengths $R_{\sigma}=7.5\kpc$ (blue dashed line) and
$R_{\sigma}=5\kpc$ (red solid line). The shorter scale-length provides the
better fit at low $v_\phi$ and the worse fit to the high-velocity tail. The
shorter length scale also yields the smaller value of the velocity-dispersion
parameter, $\sigma_0 = 22.90 \pm 0.45 \kms$ versus $\sigma_0= 24.57 \pm 0.48
\kms$.  In the plane the core of the velocity distribution is dominated by
the youngest part of the thin disc, while the wings of the distribution will
be dominated by the thick disc, and as we proceed from the core to the wings
of the distribution stars of ever increasing age will grow in importance.
Hence the true distribution in $v_\phi$ reflects the entire star-formation
history of the Galaxy and we cannot expect to obtain a perfect fit to it by adjusting a single
velocity-dispersion parameter, $\sigma_0$. Moreover, the statistics of the GCS
catalogue to some extent reflect the complex selection biases involved in the
catalogue's formation, and we have made no attempt to replicate these biases. 
Neglect of these complexities is presumably why our fit is less good than that obtained by \cite{SBD}. The
presence of kinematically hotter objects is confirmed by the estimated value
for $\sigma_0$ drifting to higher values when we expand the velocity interval
on which we perform the fits. A couple of thin-disc scale-heights above the
plane the population will be more homogeneous, being dominated by the thick
disc, and it should be possible to obtain better fits by adjusting
$\sigma_0$.

Interestingly, on testing for a systematic shift in $v_\phi$ the formula
recovered to $2 \kms$ the expected local standard of rest both from the GCS
data and from the velocity distributions of the torus models at low
altitudes. At higher altitudes the performance deteriorates due to the higher
uncertainties. We found that for local stars $\Delta E_z$ does not play a
significant role, although it gives a bias of order $1 \kms$. In a
forthcoming re-determination of the local standard of rest we will take this
effect into account.

\begin{figure}
\epsfig{file=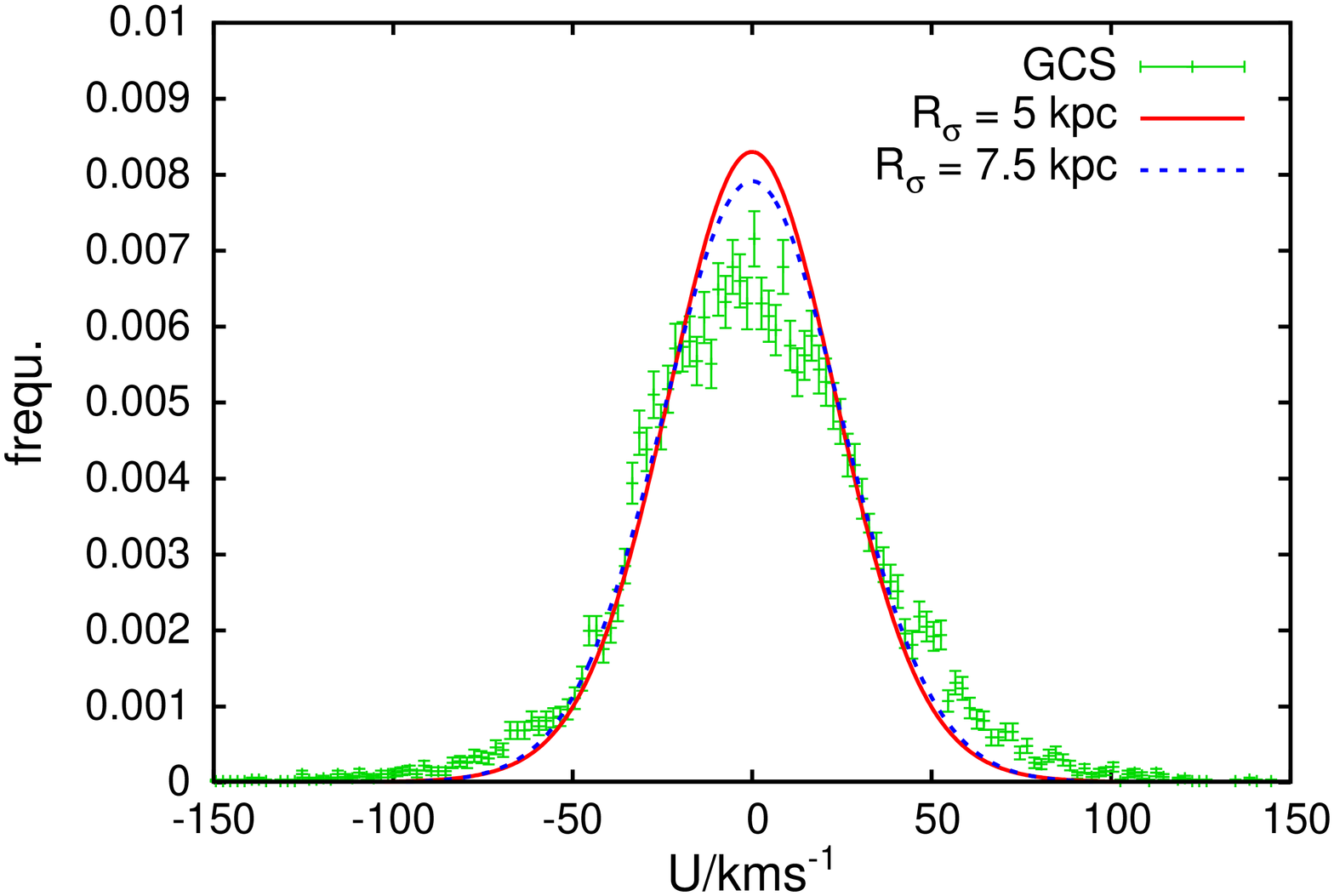,angle=-90,width=\hsize}
\epsfig{file=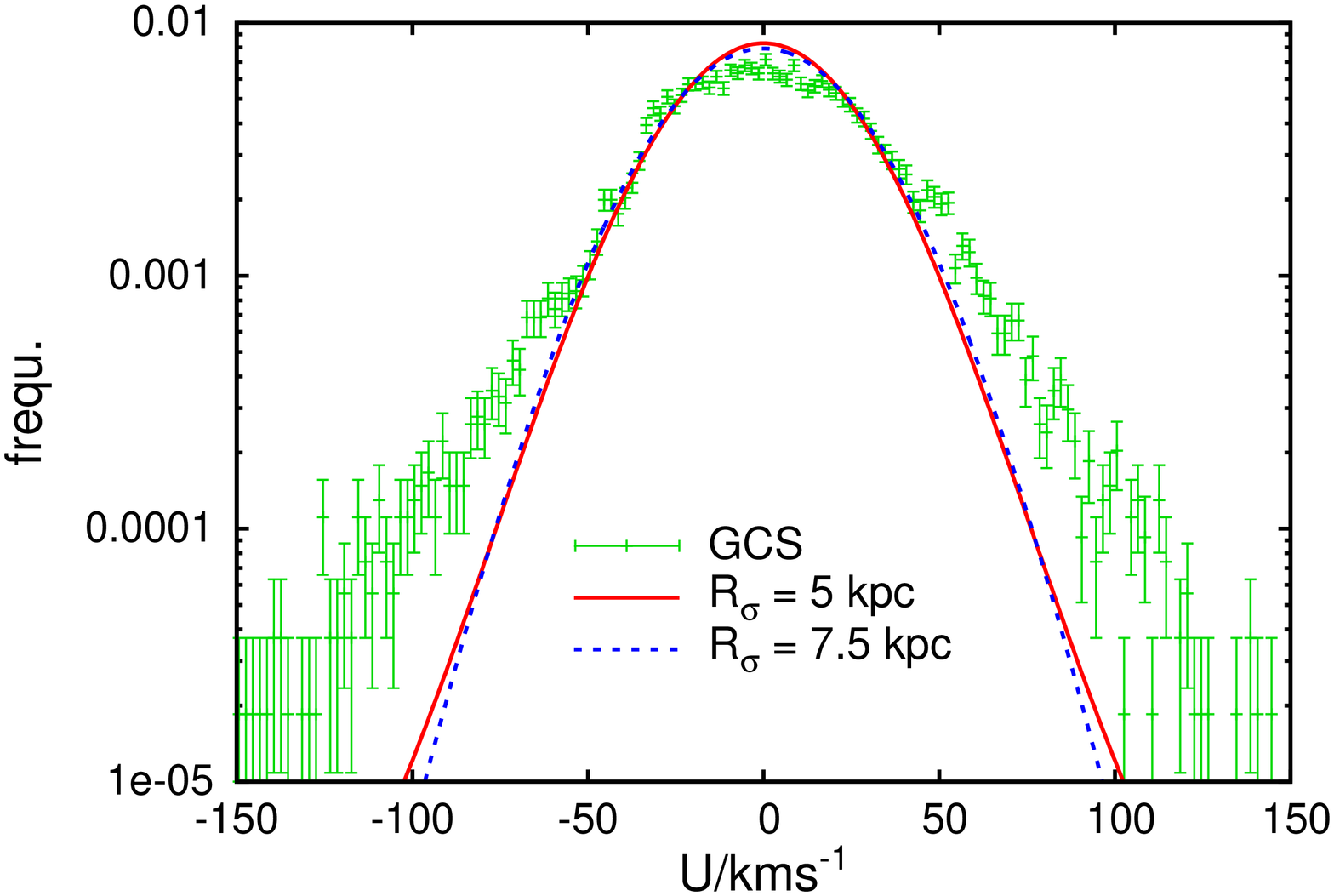,angle=-90,width=\hsize}
\caption{Derived $v_R$ velocity distributions at the parameters of Fig.\ref{fig:GCSfit} versus data.
}\label{fig:GCSfitU}
\end{figure}

Fig.~\ref{fig:GCSfitU} compares the GCS data with the $v_R$ distributions
that follow from the fits to $v_\phi$. It is clear that the formula
under-estimates the width of the $v_R$ distribution. In Section \ref{sec:VR}
we found that in the case of the torus model at small $|z|$, the fitted $v_R$
distribution was somewhat narrower than the true one even though the $v_\phi$
distribution was very closely fitted.  Correspondingly, in the case of the
GCS data, the failure of the fit to the $v_\phi$ to adequately populate
populate the wings of the distribution is mirrored in the fit to the $v_R$
distribution in Fig.~\ref{fig:GCSfitU} being least satisfactory in the wings.

\section{Conclusions}\label{sec:sum}

The distribution of azimuthal velocities in the disc of a galaxy like ours is
very skew and varies systematically with distance from the plane. Naturally
one wants to be able to quantify such a distribution in an effective way.
The traditional approach of fitting it with a superposition of Gaussians
\citep[e.g.][]{Bensby03,Ivz08,McCo06} is unsatisfactory, both because there is no
physical reasoning behind the use of a Gaussian when the distribution is not
dominated by measurement error, and because when a superposition of
Gaussians is used, the parameters of the fit are neither unique nor
physically informative.

Our formula is based on the approximation that the vertical actions of stars
are invariant as stars oscillate radially. We have refined this approximation
by considering anew the impact that vertical motion has on the radial
oscillations, which BM11 found to be a significant effect. Our treatment of
this effect, being based on overall energy conservation, is conceptually much
sounder than that of BM11 and promises to play a valuable role in the
interpretation of stellar velocities with rigorous dynamical models.
However, we find that the power of our formula is only marginally improved by
our more rigorous treatment of how vertical motion affects the radial
oscillations. 

Ultimately, our formula is just a fitting formula rather than a
dynamical theory, even though we have derived it from dynamical
considerations. Something the derivation highlights is how closely the
horizontal and vertical motions of stars are intertwined, notwithstanding the
adiabatic invariance of actions. Because both the vertical and horizontal
random velocities of stars increase with age, as one moves away from the
plane the mix of stars one sees fundamentally changes in the sense of
increasing age and decreasing radius of birth. The cleanest way to model this
phenomenon is by means of a \df\ like those presented by \cite{B10}, but a
couple of computationally challenging steps are required to extract
observationally testable velocity distribution such as $n(v_\phi)$ from a
\df: first a connection has to be established between ordinary phase-space
coordinates and the isolating integrals upon which the \df\ depends, and then
one has to marginalise over two velocities. Evaluation of our formula is
trivial by comparison.

Our derivation makes it plain that no population of stars can simultaneously
have  scale-height and velocity dispersion that are both independent
of radius:  the sub-population formed by stars that have a narrow range of
angular momenta must inevitably increase in scale-height and decrease in
vertical velocity dispersion with increasing $R$, and if stars of
larger angular momenta are added in to hold constant the scale-height, they
will have to have an even smaller vertical velocity dispersion, so the
vertical dispersion of the entire population will decline steeply outwards.
Given this situation, it is unwise to seek to define the thick disc in terms
of a given scale-height and velocity dispersion, as some recent papers have
done.

We validated our fitting formula by using it to fit the distributions of
$v_\phi$ components at several distances from the plane in a model with a
well-defined \df\ that included both thin and thick discs. Excellent fits
were obtained.  The values of the fitting parameters varied slightly with the
level of sophistication of the model employed, but were broadly in agreement
with the values we would expect given the underlying \df, especially when the
most sophisticated approximations were used. This exercise implies that
physical significance can be attached to the values of parameters derived
from fits to real data. Each fit to a $v_\phi$ distribution implies a model
of the corresponding $v_R$ distribution. In our tests these models turned out
to be very useful although showing a slight tendency to be too narrow at
small $|z|$.

We fitted the formula to the $v_\phi$ velocities of GCS stars and obtained
good but not perfect fits for plausible values of the parameters.  The
blemishes in these fits will arise from three causes: (i) the well
known presence of pronounced clumping of stars in the $(U,V)$ plane
\citep{Dehnen98}, (ii) the need to model subtle selection effects in the
GCS sample, and (iii) our formula is derived from an isothermal \df\ and
must encounter difficulty fitting data drawn from a system that is a
superposition of systems with very disparate dynamical temperatures. Hence in
part the difficulties encountered in fitting the GCS data may reflect the
importance at the extremes of the $v_\phi$ distribution of the thick disc
and/or stellar halo. With a larger body of data, or data taken further from
the plane, it might be profitable to fit the data to a sum of two or
more instances of our formula. Our fit to the $v_\phi$ components yields a
model of the $v_R$ components that is rather too narrow, in agreement with
our work with the model based on a \df.

Our formula could be used to fit the line-of-sight velocity distributions
(LOSVDs) of galaxies in which individual stars are not resolved
\citep[e.g.][]{sauron}. Since our formula has been derived on the assumption
that the circular speed is independent of radius, it might fail to produce a
satisfactory fit to the velocity distribution of stars in a galaxy with a
distinctly non-flat rotation curve. The physical basis of our formula breaks
down at distances from the plane of order $2\kpc$, so failure to fit data for
stars at higher altitudes might have no physical significance.

We derived our formula by adapting to the three-dimensional world the planar
\df\ of \cite{Shu69}. It would be interesting to adapt in a similar way the
planar \df\ of \cite{Dehnen99}, which Dehnen has argued is in certain
respects superior to the Shu \df. Unfortunately, the adaptation of a
planar \df\ is a non-trivial exercise on account of the intertwining of the radial and
vertical motions mentioned above. Therefore in this paper we have confined
ourselves to the Shu \df, which proves to provide a very useful point of
departure.

Data for stars that lie at significant distances from the plane are now
becoming available \citep{Ivz08,RAVE3}. The measured space velocities of such
stars contain significant errors arising from a combination of errors in
proper-motion and distance. It is essential to take proper account of these
errors when inferring the true kinematics of the underlying populations. Our
formula provides the natural way to do this: one fits the data to the result
of folding the formula with appropriate distance and proper-motion errors.
\cite{SAC} use this methodology to extract in an elegant way the
information contained within the measured distribution of azimuthal
velocities of stars that have quite large random velocities. We will shortly
present similar analyses of samples that include a higher proportion of disc
stars.

\section*{Acknowledgements}
It is a pleasure to thank Andreas Just for detailed reading of the paper and many very valuable comments.
We thank Walter Dehnen for helpful comments on a draft and Paul McMillan for the compilation and discussion of the torus model data. R.S. acknowledges financial and material support from
Max-Planck-Gesellschaft. J.B. acknowledges the support of STFC and Merton
College, Oxford.

\section*{Appendix:  estimating  the typical vertical energy}

In this appendix we estimate the typical value $\overline{E_z}(R_g)$ of the vertical energy
$E_z(J_z,R)$ of the stars that we encounter with specified angular momentum
$L_z=Rv_\phi$ at a given location $(R,z)$. Let $\sigma_z$ be the vertical velocity
dispersion at $(R,z)$. then 
 \begin{equation}
\overline{E_z}(R_g)\simeq\fracj12\sigma_z^2(R_g)+\Phi_z(z).
\end{equation}
 Regarding $\sigma_z$, we have that to an excellent approximation the
vertical Jeans equation reads \citep[][eq.~4.271]{BT08}
 \begin{equation}
{\d\left(\rho\sigma_z^2\right)\over\d z}=-\rho{\d\Phi_z\over\d z}.
\end{equation}
 Neglecting the derivative of $\ln\sigma_z^2$ relative to that of $\ln\rho$
this yields 
\begin{equation}
\sigma_z^2=h{\d\Phi_z\over\d z},
\end{equation}
 where $h\equiv-(\d\ln\rho/\d z)^{-1}$ is the local scale-height of the
population. As we saw from equation (\ref{eq:giveshrr}) the local scale-height
decreases towards lower guiding centre radii, if we assume all populations to
have at their guiding centre radius a constant scale-height. Hence finally we
adopt:
 \begin{equation}\label{eq:Ezbar}
\overline{E_z}(R_g)=\fracj12h(R_g,R){\d\Phi_z\over\d z}\bigg|_z+\Phi_z(z).
\end{equation}

Equation (\ref{eq:Ezbar})
involves the  potential and its derivative at altitude $z$.
We obtain these from a simple mass model with a razor
thin gas layer that at the solar radius has surface density
$12 \msun \pc^{-2}$ and three exponential stellar components. In each such
component the mass per unit surface area within distance $z$ of the plane is
\begin{equation}
\Sigma_i(z) = \Sigma_{i,0}\left(1-\e^{-z/h_i} \right).
\end{equation}
 The three components represent the thin disc, the thick disc and the halo
with parameters
\begin{equation} 
(\Sigma_{1, 0}, \Sigma_{2, 0}, \Sigma_{3, 0}) = (30, 10, 70) \msun \pc^{-2}
\end{equation}
and 
\begin{equation}
(h_1, h_2, h_3) = (300, 1000, 4000) \pc.
\end{equation}
 The halo contribution was chosen to match the local vertical potential from
the adopted Dehnen potential, which was used for the torus models to which we
compare our formalism -- the modifications required for a different radius or
disc mass are simple. We approximate the
contribution $\Phi_i(z)$ to the potential from the $i$th component by
assuming that the component is an infinite plane-parallel sheet. Then
\begin{equation}
\Phi_i(z) = 2\pi G \Sigma_{i, 0} \left[z + h_i\left(\e^{-z/h_i} - 1
\right)\right].
\end{equation}

\label{lastpage}
\end{document}